\documentclass[aps,twocolumn,nobalancelastpage,amsmath,amssymb,
    nofootinbib,longbibliography]{revtex4}
\usepackage[utf8]{inputenc}

\usepackage{hyperref} 
\hypersetup{
    pdftitle={Berry Phase},
    unicode=false,          
    pdftoolbar=true,        
    pdfmenubar=true,        
    pdffitwindow=false,     
    pdfstartview={FitH},    
    pdfsubject={},   
    pdfcreator={},   
    pdfproducer={}, 
    pdfkeywords={} {} {}, 
    pdfnewwindow=true,      
    colorlinks=true,       
    linkcolor=magenta, 
    citecolor=blue,        
    filecolor=magenta,      
    urlcolor=blue           
} 

\usepackage{mathtools}
\usepackage{diagbox}
\usepackage{savesym}
\usepackage{wasysym}
\usepackage{amscd}
\usepackage{graphicx}
\usepackage{pifont}
\usepackage{float}
\usepackage[all]{xy}
\usepackage{multirow}
\usepackage{amsfonts}
\usepackage{picture}
\usepackage{amssymb}
\savesymbol{iint}
\savesymbol{iiint}
\usepackage{amsmath}
\restoresymbol{TXF}{iint}
\restoresymbol{TXF}{iiint}
\usepackage{color}

\usepackage{graphicx}
\usepackage{dcolumn}
\usepackage{tikz}
\usetikzlibrary{calc,decorations.pathmorphing,arrows,decorations.markings,cd}
\tikzset{snake it/.style={decorate, decoration=snake}}
\usetikzlibrary{shapes}
\usetikzlibrary{patterns}
\usetikzlibrary{arrows,positioning,decorations.pathmorphing,
  decorations.markings, matrix, decorations.text}

\usepackage{setspace}
\usepackage{multirow}
\usepackage{wrapfig}
\usepackage{verbatim}
\usepackage{dsfont}
\usepackage{slashed}
\usepackage[vcentermath]{youngtab}

\numberwithin{equation}{section}
\usepackage{float}
\restylefloat{table}
\allowdisplaybreaks
\usepackage{accents}
\usepackage{bbm}

\def\tilde{\widetilde}

\def\hat{\widehat}

\def\bar{\overline}

\def\eps{\epsilon}
\def\1{{\mathds 1}}
\def\ra{\rightarrow}

\def\CD{{\mathcal D}}

\def\CK{{\mathcal K}}
\def\CL{{\mathcal L}}
\def\CM{{\mathfrak M}}

\def\CP{{\mathcal P}}

\def\CT{{\mathcal T}}

\newcommand{\ZZZ}{{\mathbb Z}}
\newcommand{\RRR}{{\mathbb R}}
\newcommand{\PPP}{{\mathbb P}}

\newcommand{\CMU}{{\mathfrak M}^{U(1)}}

\usepackage{datetime}

\begin{document}
\title{\Large Berry Phase in Quantum Field Theory: \\ \large Diabolical Points and Boundary Phenomena}

\author{Po-Shen Hsin$^1$}
\author{Anton Kapustin$^1$}
\author{Ryan Thorngren$^2$}
\affiliation{$^1$Walter Burke Institute for Theoretical Physics, California Institute of Technology, Pasadena, CA 91125}
\affiliation{$^2$Center of Mathematical Sciences and Applications, Harvard University, Cambridge, MA 02138}

\date{\today}
\begin{abstract}
We study aspects of Berry phase in gapped many-body quantum systems by means of effective field theory. Once the parameters are promoted to spacetime-dependent background fields, such adiabatic phases are described by  Wess-Zumino-Witten (WZW) and similar terms.  In the presence of symmetries, there are also quantized invariants capturing generalized Thouless pumps. Consideration of these terms provides constraints on the phase diagram of many-body systems, implying the existence of gapless points in the phase diagram which are stable for topological reasons. We describe such diabolical points, realized by free fermions and gauge theories in various dimensions, which act as sources of ``higher Berry curvature" and are protected by the quantization of the corresponding WZW terms or Thouless pump terms. These are analogous to Weyl nodes in a semimetal band structure. We argue that in the presence of a boundary, there are boundary diabolical points---parameter values where the boundary gap closes---which occupy arcs ending at the bulk diabolical points. Thus the boundary has an ``anomaly in the space of couplings" in the sense of \cite{Cordovaetal1,Cordovaetal2}. Consideration of the topological effective action for the parameters also provides some new checks on conjectured infrared dualities and deconfined quantum criticality in 2+1d.
\end{abstract}

\maketitle

\section{Introduction}

In this article, we explore topological invariants associated with the global  structure of phase diagrams of quantum systems at zero temperature and their physical consequences. An important source of topology in phase diagrams is the presence of so-called ``diabolical points" \cite{BerryDiabolical} (or more general ``diabolical loci")---isolated parameter values where the ground state differs from the surrounding phase, for example by being degenerate or gapless.

A familiar example is a stationary spin-1/2 in a magnetic field. At zero magnetic field, the energy spectrum is doubly degenerate, while at any non-zero magnetic field there is a unique ground state. One has to tune three real parameters, namely the components of the magnetic field, to encounter this degeneracy. A classic result of von Neumann and Wigner \cite{vN_Wigner} states that this is a generic situation: in quantum systems with a finite number of degrees of freedom  and without symmetries\footnote{When a unitary symmetry is present, energy levels  which transform in different representations of the symmetry group generically cross in codimension-1, since there is no level repulsion in this case.} energy level degeneracy generically occurs in codimension 3.  

Another classic observation, due to M. V. Berry \cite{Berry}, is that such a ``diabolical locus" $\CD$ can be detected by studying the ground-state wave-function on the complement of $\CD$.  Consider the case when the parameter space is three-dimensional and $\CD$ is a single point. Removing $\CD$ from its small neighborhood we end up with a space which is homotopy equivalent to a two-dimensional sphere. One can show that because of a non-integrable Berry phase a properly normalized ground-state wave-function depending continuously on parameters can not be defined globally on this two-sphere. The best one can do is to let the ground-state be a normalized continuous section of a nontrivial line bundle over the two-sphere \cite{Simon}.
A manifestation of this is that the curvature of the Berry connection has a monopole singularity at $\CD$.

In many-body quantum systems, degeneracy of the ground state generically occurs in codimension $1$ rather than $3$. These are loci where the system undergoes a 1st order phase transition. The von Neumann-Wigner result does not apply to such situations, and the degeneracy locus does not leave a topological imprint. Nevertheless, many-body analogues of diabolical points do exist. These are gapless points or gapless loci in the phase diagram which are entirely surrounded by the trivial phase (although they can also occur inside other phases). Such gapless points are not simply endpoints of 1st order transitions, they are robust for topological reasons. The simplest example is a free Dirac fermion in spatial dimension $d$ with charge conservation. Depending on whether $d$ is even or odd, the gapless point in this system occurs in codimension $1$ or $2$. Usually its stability is explained using anomalies (parity anomaly for even $d$ and axial anomaly for odd $d$). But there also exist examples of isolated gapless points which cannot be explained using anomalies. We propose a framework which treats all these examples on the same footing. 

A natural generalization of the Berry phase to $d>0$ is the Wess-Zumino-Witten (WZW) term in the effective action for the parameters. A WZW term in spatial dimension $d$ is described by a closed $(d+2)$-form on the parameter space which may have monopole-like singularities along gapless loci. Such singularities occur in codimension $d+3$, therefore one might conclude that diabolical points in $d$ dimensions occur in codimension $d+3$. In fact, the situation is more complex, and the WZW form describes just a particular class of topological terms in the effective action. In general, there are several types of diabolical points in a given dimension, and their classification also depends on whether the system is bosonic or fermionic. Imposing symmetry also affects the classification of diabolical points. This was pointed out already by von Neumann and Wigner who showed that in the presence of time-reversal symmetry diabolical points in 0+1d systems occur in codimension 2  \cite{vN_Wigner}. In any case, using the effective action approach one can classify possible diabolical loci and prove their stability.

Another point of view on this problem, which is very powerful, comes from considering the infinite-dimensional space of all field theories, {\it i.e.} UV-complete theories defined at some UV scale. This space is decomposed into strata defined by the attractive basins of the RG flow, consisting of field theories which flow to a particular RG fixed point in the IR. Roughly speaking, the number of relevant operators at the fixed point gives the codimension of the stratum.\footnote{This is complicated by theories with marginally relevant or dangerously irrelevant operators, which demonstrate that this stratification is quite singular \cite{Gukov_2016}.} A generic phase diagram with $k$ parameters is a $k$-dimensional slice through this infinite dimensional space, transverse to all the strata, meaning a stratum with $k$ relevant operators at the fixed point occurs at isolated points in this $k$-parameter phase diagram. Protected diabolical loci correspond to the situation where one stratum pokes a hole in another, leaving the latter with some nontrivial topology. This topology, in turn allows for the universal definition of topological terms in the effective action, associated with a certain cohomology of this stratum. Thus, the higher Berry phase and protected diabolical loci points of view are equivalent.


An important novelty of working in higher dimensions is the possibility of studying spatial boundaries. We will describe a boundary-bulk correspondence for these ``higher" Berry phases, which is reminiscent of the Fermi arcs in Weyl semimetals \cite{weylsemimetal11}. Indeed, we observe that relative to some choice of boundary condition, the higher Berry phase implies the existence of a diabolical locus in parameter space where the boundary gap closes, whose boundary in parameter space is in the bulk diabolical locus. In the context of lattice models, this observation was recently made in \cite{KapSpo1,KapSpo2}. We also make contact with the ``anomalies in the space of coupling constants" studied recently by \cite{Cordovaetal1,Cordovaetal2}.

Many of the physical phenomena associated with higher Berry phase are well-known and can be exhibited by free fermion systems. See \cite{TeoKane} for a general approach in this case. A particularly useful collection of such models was defined by Abanov and Wiegmann \cite{AbanovWiegmann}. As we review in Appendix \ref{appdimredAW}, they give rise to $S^{d+2}$ level 1 WZW terms and $S^d$ generalized Thouless pumps. We will discuss several of these theories in some detail, including their boundary phenomena, as well as a free fermion model in 3+1d ``missing" from \cite{AbanovWiegmann}, which is associated with the $p+ip$ superconductor/gravitational Chern-Simons term. 

While free fermion theories provide many examples of diabolical points, we stress that our arguments prove their stability on the non-perturbative level. We also discuss some interacting models (gauge theories). 


The paper is organized as follows. In Section \ref{sec1+1d} we discuss free fermion diabolical points and their corresponding families in 1+1d. There is a codimension 2 diabolical point protected by the Thouless pump and a codimension 4 diabolical point protected by a higher Berry number. We discuss the physics of interfaces and boundaries for both families.

In Section \ref{sec2+1d} we extend the discussion to free fermion diabolical points and their families in 2+1d. There are again two types, corresponding to a family with a charged skyrmion, analogous to the Thouless pump in 1+1d, and a family with a higher Berry number, respectively.

In Section \ref{sec3+1d} we discuss free fermion diabolical points and their families in 3+1d. There is a new codimension 2 diabolical point in this dimension which generates an axion coupling, and corresponds to a Thouless pump of a gravitational Chern-Simons term.

In Section \ref{secgaugetheories} we discuss interacting examples of diabolical points and their families in gauge theories. This includes 2+1d Chern-Simons matter theories realizing the generalized Thouless pump. The nontrivial families are consistent with the recently studied web of 2+1d IR dualities, and we interpret the deconfined quantum critical point between the N{\'e}el state and the valence bond solid (VBS) as a protected diabolical point. We also discuss applications to scalar QED in 3+1d.

In Section \ref{secbulkboundary}, we provide a general proof of the bulk-boundary correspondence for the higher Berry phase and Thouless pumps. The result describes an index theorem whereby there will be topologically protected boundary diabolical points, the sum of whose higher Berry numbers equals the bulk higher Berry number.

In Section \ref{secgenremarks}, we discuss some mathematical formalism for describing families of gapped phases, including a classification based on cobordism theory and TQFT.

\section{Free fermions in 1+1d}\label{sec1+1d}

\subsection{The Thouless pump and a codimension-2 diabolical point}\label{subseccomplexferms1d}

\subsubsection{A topologically nontrivial family}

As a motivating example, consider a single complex fermion in 1+1d with a Minkowski-signature Lagrangian
\begin{equation}\label{AWmodelA1d}
\CL_{Dirac}=-\bar\psi i\slashed{\partial}\psi - i\bar\psi(M_1+iM_2\gamma^{01})\psi.
\end{equation}
Here $\slashed{\partial}=\gamma^\mu\partial_\mu$, $\gamma^0$ and $\gamma^1$ are 2d Dirac matrices satisfying $\{\gamma^\mu,\gamma^\nu\}=2g^{\mu\nu}$, $g^{\mu\nu}={\rm diag}(-1,1)$ is the Minkowski metric, and $\gamma^{01}=\gamma^0\gamma^1$ is the chirality operator\footnote{For definiteness, in terms of the usual Pauli spin matrices, one can choose $\gamma^0 = i\sigma^y$, $\gamma^1 = \sigma^x$, $\gamma^{01} = \sigma^z$.}. Some unusual-looking signs and factors of $i$ in (\ref{AWmodelA1d}) and similar formulas below are due to our use of ``mostly plus" convention for the Minkowski metric. Our conventions are chosen to facilitate comparison with Ref. \cite{AbanovWiegmann}. 

Hermiticity of the Hamiltonian requires $M_1$ and $M_2$ to be real, and it is convenient to introduce a complex mass $M=M_1+i M_2$. We will also denote $\alpha={\rm arg}\, M.$ This Lagrangian has a $U(1)$ symmetry under which $\psi$ has charge $1$ and $\bar\psi=\psi^\dagger \gamma^0$ has charge $-1$.  The axial $U(1)$ transformation $\psi\mapsto e^{i\alpha\gamma^{01}}\psi$ is not a symmetry except for $M=0$, since it has the effect of replacing $M$ with $M e^{2i\alpha}$. For $M\neq 0$ this model describes massive fermionic particles and anti-particles of mass $|M|$, while for $M=0$ it is gapless. Its properties for spatially-varying $M$ have been studied by Goldstone and Wilczek \cite{GoldstoneWilczek}. It is also a special case of the ``A" series of Abanov and Wiegmann  \cite{AbanovWiegmann}.

In this model the massless point is an isolated gapless point of codimension $2$. One may ask if it is possible to gap it out by adding interactions or couplings to other massive degrees of freedom, without breaking $U(1)$ or bringing in new gapless points from infinity. The answer is no, and one way to see it is to couple the theory for $|M| > 0$ to a background $U(1)$ gauge field and study the effective action after we integrate out the fermion. We will find a topological term whose presence is inconsistent with a completely gapped phase diagram.

From the axial anomaly, one finds that this effective action contains a term
\begin{equation}
\frac{1}{2\pi}\int \alpha\,  \eps^{\mu\nu} F_{\mu\nu} d^2 x=\frac{1}{2\pi}\int \alpha\,F
\end{equation}
where $F=dA=\frac12 F_{\mu\nu} dx^\mu dx^\nu$ is the gauge field strength. This term does not seem to have much effect if we regard $A$ as a background gauge field and if $M$ is a constant. But as noticed by Goldstone and Wilczek \cite{GoldstoneWilczek}, the situation changes if we allow $M$ to be a slowly-varying function of coordinates. The above action is then the leading term in the derivative expansion in $M$ and gives rise to a topological contribution to the $U(1)$ current:
\begin{equation}\label{Jdirac}
J^\mu_\text{top}=\frac{1}{2\pi}\eps^{\mu\nu}\partial_\nu \alpha.
\end{equation}
This $U(1)$ current represents the response of the fermionic vacuum to an adiabatic variation of $M$. If $M$ depends only on $x^1$, it gives a $U(1)$ charge density to the vacuum. The net charge of a configuration where $\alpha={\arg M}$ approaches $\alpha_\pm$ as $x\ra\pm\infty$ and slowly varies in between is $(\alpha_+-\alpha_-)/2\pi+n$, where $n$ is the (signed) number of times $M$ crosses the branch cut for the logarithm. In general the charge is fractional. If $\alpha_+=\alpha_-$, this charge is just $n$ and can be interpreted as the collective charge of $n$ winding solitons in $\alpha$ \cite{GoldstoneWilczek}. 
If $M$ depends only on $x^0$, there is a uniform vacuum current. In particular, if $M$ is a periodic function of time and winds around the origin $N$ times per period, the net charge which flows through a section of a system during this time is $-N$. This phenomenon is known as the Thouless charge pump \cite{Thouless}.

The presence of this topological term implies that we cannot completely remove the gapless point at the origin in a $U(1)$-invariant way by adding UV degrees of freedom and continuously changing their couplings to the field $\psi$ (although we can deform it into a more general diabolical locus, see Appendix \ref{appluttingerthouless}). This is because for a gapped nondegenerate family on $\RRR^2$, we would need to extend the form  $\epsilon^{\mu \nu}\partial_\nu \alpha$, required to express the topological current, to maps $M$ valued anywhere in the plane, which is impossible.

To formalize this argument, we interpret the loop of models defined by $|M| = m > 0$ as a noncontractible loop in the space $\CM^{U(1)}_1$ of 1+1d field theories with a unique gapped ground state ({\it i.e.} infrared-trivial theories) and $U(1)$ symmetry.

Consider a general such theory defined over a parameter space $P$. The low energy limit defines a map from our parameter space $P \to \CMU_1$. Slowly-varying parameters in spacetime thus define a map $\phi:X\ra\CMU_1$. We can define an effective action for $\phi$. Expanding in derivatives of $\phi$, the leading term has the form
\begin{equation}\label{Thouless1d}
S_\text{top}=\int_X \eps^{\mu\nu} A_\mu \partial_\nu \phi^i \tau_{1,i}(\phi) d^2x+\ldots=\int_X A\wedge \phi^*\tau_1+\ldots,
\end{equation}
where $\tau_1=\tau_{1,i}(\phi) d\phi^i$ is a 1-form on $\CMU_1$ (that is, it is universal) and dots denote terms which are independent of both $A$ and the metric. Invariance with respect to infinitesimal gauge transformations $A\mapsto A+d f$ requires the 1-form $\tau_1$ to be closed. Invariance with respect to large gauge transformations requires it to have integral periods. That is, for any loop $\phi:S^1\ra\CMU_1$ one must have $\oint \phi^*\tau_1\in\ZZZ$. This integer is associated with the Thouless pump of the loop. Indeed, the topological current corresponding to the above action is obtained by varying with respect to $A_\mu$:
\begin{equation}
J^\mu_\text{top}=\eps^{\mu\nu} \tau_{1,i}(\phi) \partial_\nu\phi^i.
\end{equation}
This current leads to the same physical consequences as the current (\ref{Jdirac}). Namely, static configurations of the field $\phi$ acquire $U(1)$ charge, while a time-dependent $\phi$ produces a spatial current. Given a loop $\phi:S^1\ra\CMU$ with a basepoint $\phi_0\in\CMU_1$, one can attach to it a solitonic configuration on $\RRR$ which approaches $\phi_0$ at $x=\pm\infty$. The charge of this soliton is equal to $\oint \tau_{1,i} d\phi^i$ and is integral. We can also identify $S^1$ with the time coordinate thus making $\phi$ a periodic function of time. The net charge transported per period is $\Delta Q=-\oint \tau_{1,i} d\phi^i$.

If a given loop in $\CMU_1$ is contractible, then the integral of $\tau_1$ over this loop vanishes. Mathematically, this follows from the Stokes theorem. The physics of this is also clear: if a slowly varying field configuration on $\RRR$ can be deformed to the constant one without changing the asymptotics at $x=\pm\infty$ and without creating large gradients, its charge must be zero.\footnote{If the gradients become large, there may be level crossing, and the vacuum charge may jump discontinuously.} In the case of the complex fermion in 1+1d, it follows from Eq. (\ref{Jdirac}) that $\tau_1=\frac{1}{2\pi} d\alpha$. Therefore the integral of $\tau_1$ over the loop in the $M$-plane circling the origin is $1$, and thus  this loop is not contractible. Any nonzero multiple of this loop is also not contractible in $\CMU_1$. Thus the fundamental group of $\CMU_1$ contains a copy of $\ZZZ$.\footnote{According to the classification we discuss in Section \ref{secgenremarks}, $\pi_1 (\CMU_1) = \ZZZ$. Thus, $\tau_1$ is a complete invariant describing the family: two families with the same $\oint \phi^*\tau_1$ may be deformed into each other without closing the gap.} A more precise way to phrase it is this. Let $\CM^{Dirac}_1$ denote the subspace of $\CMU_1$ corresponding to the family of theories (\ref{AWmodelA1d}) with $M\neq 0$. Obviously, $\CM^{Dirac}_1$ is the $M$-plane with the origin removed and thus $\pi_1(\CM^{Dirac}_1)=\ZZZ$.  The above argument shows that the map $\pi_1(\CM^{Dirac}_1)\ra \pi_1(\CMU_1)$ arising from the embedding $\CM^{Dirac}_1\ra \CMU_1$ is injective. This implies that the isolated gapless point at $M=0$ cannot be gotten rid by arbitrary deformations which preserve $U(1)$ symmetry.

Although the diabolical point cannot be completely removed, the nature of the diabolical locus can be modified by deforming the theory for small $|M|$. For example, one can deform the free fermion theory into a more general Luttinger liquid. Depending on how this is done, one can ``resolve'' the diabolical point into a 1st order phase transition line terminating at two critical Ising points, or into an island of a gapless Berezinsky-Kosterlitz-Thouless phase. This is described in more detail in Appendix \ref{appluttingerthouless}.

\subsubsection{Interfaces}\label{subsubsec1dthoulessinterfaces}


A physical consequence of non-contractibility can be seen when one studies smooth interfaces between different models in the family. For the Dirac fermion \eqref{AWmodelA1d}, we will study interfaces where the mass parameter $M = m e^{i \alpha}$, $m = |M|$, varies from $m$ at $x = -\infty$ to some other value $m e^{i \alpha_0}$ at $x = \infty$. In particular we will be interested in 1-parameter families of interfaces parametrized by $\alpha_0 \in \RRR/2\pi\ZZZ$. We will show that for at least one of these interfaces, there must be a localized zero mode.

The simplest option is to take $m$ constant and let $\alpha(x)$ be a step-function, $\alpha(x)=\alpha_0\theta(x)$. This gives a family of interfaces which is periodic as a function of $\alpha_0$ but is not smooth in $x$-space. While this is not quite what we need, the advantage of this family of interfaces is that it is soluble \cite{MacKenzie:1984mz} and one can find the states localized on the interface exactly. One lets
$\psi=\psi(0)e^{-a x}$ for $x>0$ and $\psi=\psi(0)e^{ax}$ for $x<0$, with some $a>0$. Then  $\psi(0)$ satisfies
\begin{align}
&\left(\begin{array}{cc}
-ia & me^{i\alpha_0}\\
me^{-i\alpha_0}& ia
\end{array}
\right)
\psi(0)=E\psi(0),\cr
&\left(\begin{array}{cc}
ia & m\\
m& -ia
\end{array}
\right)
\psi(0)=E\psi(0)~.
\end{align}
This gives
\begin{equation}
a=m\sin\frac{\alpha_0}{2},\quad E=m\cos\frac{\alpha_0}{2}~.
\end{equation}
For $\alpha_0\in (0,2\pi)$ there is a single mode localized on the interface. Its energy becomes zero at $\alpha_0=\pi$. The mode becomes non-normalizable and merges with the continuum for $\alpha_0=0$ (or $\alpha_0=2\pi$).

Another soluble case is an interface where $\alpha(x)$ is continuous, varies linearly with $x$ in an interval $0<x<L$, and is equal to its asymptotic values $0$ and $\alpha_0$ for $x\leq 0$ and $x\geq L$ respectively. This interface was analyzed in \cite{KeilKobes}.
One finds that for small and positive $\alpha_0$ there is a single mode on the interface whose energy is just below $m$ and decreases monotonically as $\alpha_0$ increases. The energy becomes zero for $\alpha_0$ of order $mL$. At certain threshold values of $\alpha_0$ additional interface modes appear. This family of interfaces is not periodic in $\alpha_0$, {\it i.e.} the interface at $\alpha_0=2\pi$ is not identical to the interface at $\alpha_0=0$.

To get a family of interfaces which is smooth and periodic in $\alpha_0$, one needs to make both $|M|$ and $\alpha={\rm arg}\, M$ dependent on $x$. For example, one can take $M(x)$ to take values in
a straight-line segment connecting $m$ and $m e^{i\alpha_0}$. This family of interfaces does not appear to be soluble. Nevertheless, its qualitative behavior is easy to understand. For $\alpha_0\ll 2\pi$ this family of interfaces coincides with that studied in \cite{KeilKobes}, up to terms which are quadratic in $\alpha_0$. Thus there is a single normalizable mode on the interface whose energy is slightly below $m$ and approaches $m$ in the limit $\alpha_0\ra 0$.

As one increases $\alpha_0$, additional interface modes may appear. Every time such a mode crosses zero energy from above (resp. from below), the ground-state charge for the effective quantum mechanics describing the interface jumps by $+1$ (resp. $-1$). It is easy to show that the algebraic number of such crossings ({\it i.e.} the number of crossings from above minus the number of crossings from below) as $\alpha_0$ increases from $0$ to $2\pi$ is either $1$ or $-1$. Indeed, charge conjugation symmetry tells us that each crossing at $\alpha_0=b$ is accompanied by a crossing at $\alpha_0=2\pi-b$, and their contributions are opposite.

The only unpaired crossing may occur at $\alpha_0=\pi$. This situation corresponds to $M(x)$ being real for all $x$ and varying from $m$ to $-m$. It was shown by Jackiw and Rebbi \cite{JackiwRebbi} that there is exactly one  normalizable zero mode in this case. Thus the algebraic number of crossings is $\pm 1$. One can verify that it is $1$ by taking the limit $L\ra 0$. Then we get an infinitely thin interface studied by MacKenzie and Wilczek \cite{MacKenzie:1984mz}, and we know that in that case the zero energy level is crossed from above. 

To understand the fate of general interfaces, even those which do not come simply as a spatially-varying mass profile, we can appeal to the topology of $\CMU_1$. An infrared-trivial interface can be thought of as a path in $\CMU_1$. Let us fix the theory at $x = -\infty$ (the start of the path) to be the Dirac fermion with $M = m > 0$, and we define the space $\CP_{m}\CMU_1$ of paths in $\CMU_1$ beginning at this point. There is a map $r:\CP_{m}\CMU_1 \to \CMU_1$ which is given by the theory at large $x$. A periodic, smooth 1-parameter family of infrared-trivial interfaces between the Dirac fermion with $M = m > 0$ and the Dirac fermion with $M = m e^{i\alpha_0}$ at $x = \infty$ is the same thing as a map $f:S^1_{\alpha_0} \to \CP_m\CMU_1$, such that $r \circ f: S^1_{\alpha_0} \to \CMU_1$ is our Dirac family with a  spatially-constant mass $M = me^{i\alpha_0}$. However, this is impossible, because that family is noncontractible in $\CMU_1$, but $\CP_m \CMU_1$ is a contractible space: all paths can be continuously retracted to the constant path at $m$. This is a contradiction, so for any periodic smooth family of interfaces with these end points, there must be at least one interface with a degenerate ground state.

\subsubsection{Boundary-bulk correspondence}

We may regard such behavior as an anomaly in the space of couplings for the interface, in the sense of \cite{Cordovaetal1}. Imagine integrating out all the bulk modes, leaving only the modes bound to the interface. This gives an effective 0+1d field theory (that is, quantum mechanics) with a $U(1)$ symmetry. For a generic value of $\alpha_0$ the ground state of this 0+1d field theory is unique and has a well-defined charge. The only exceptions are points in the parameter space where the energy of some modes becomes zero. Every time the energy of a mode crosses zero from above (resp. from below), the ground-state charge increases by $1$ (resp. decreases by $1$). In view of the above discussion, as one increases $\alpha_0$ from $0$ to $2\pi$ the net change of the ground state charge is $1$. 
 
Such a situation would be impossible if there were a UV-complete effective quantum mechanics for the interface modes for all $\alpha_0$. Indeed, we would expect a well-defined effective action away from the gapless points. This effective action is 
\begin{equation}\label{eqnboundarycharge}
S_{int}=\int q(\alpha_0) A,
\end{equation}
where $q:S^1\ra\ZZZ$ is a locally-constant continuous function (the vacuum charge) and $A$ is a background $U(1)$ gauge field restricted to the interface. The function $q$ can change only at gapless points. Since $q$ is single-valued, the sum of jumps of $q$ across all gapless points is obviously zero. This means that the situation described in the previous paragraph cannot occur in a family of UV-complete 0+1d theories parameterized by $S^1$. However, it can and does occur in the presence of a 1+1d bulk. The physical reason for this is that at certain points in the parameter space the interface modes become non-normalizable and merge with the bulk excitations. The function $q$ can change discontinuously at these ``delocalization" points as well as at gapless points. The sum of jumps over both types of special points vanishes. 

One can quantify the anomaly of the boundary 0+1d field theory as the sum of jumps of $q$ at the gapless points, or equivalently as minus the sum of jumps of $q$ over ``delocalization" points. The latter viewpoint makes apparent the relation of the anomaly with the Thouless pump invariant of the bulk system: the net charge which flows through the system as one cycles in the parameter space should be equal to the net charge of the interface modes which became delocalized and escaped to infinity. This gives us a ``boundary-bulk correspondence": the sum of jumps of $q(\alpha_0)$ over all gapless points is equal to the Thouless pump invariant of the bulk. This may also be reasoned from the topological terms \eqref{eqnboundarycharge}, \eqref{Thouless1d}. We will revisit this correspondence more generally in Section \ref{secbulkboundary}.


\subsubsection{Reducing \texorpdfstring{$U(1)$}{U(1)} to fermion parity and torsion}\label{subsubsecparitypump}

It is clear from the above discussions that the nontriviality of this family rests on the conservation of $U(1)$ charge. We can consider reducing the symmetry so that only the $\ZZZ_2$ fermion parity is conserved.\footnote{This system satisfies a spin charge relation, so this $\ZZZ_2$ is a subgroup of $U(1)$. Formally, this means that $A$ is not a $U(1)$ gauge field but a $Spin^c$ structure.} We expect that the Thouless pump invariants will all be reduced from $\ZZZ$ to $\ZZZ_2$. Formally, we regard our family \eqref{AWmodelA1d} for some $|M| = m > 0$  as a loop in the space of arbitrary (fermionic) infrared-trivial 1+1d field theories $\CM_1$. There is a map $\CMU_1 \to \CM_1$ given by forgetting the $U(1)$ symmetry. We will show the induced map on $\pi_1 (\CMU_1) = \ZZZ \to \pi_1 (\CM_1)$ is reduction modulo 2.


We do this in two steps. First, let us argue that the  family \eqref{AWmodelA1d} remains nontrivial after breaking $U(1)$ to $\ZZZ_2$. This is simply because by the spin-charge relation, our soliton which carries one unit of $U(1)$ charge is fermionic. On the other hand, the constant ``loop" ({\it i.e.} with constant $M$) corresponds to a state with an even fermion parity. This means we will continue to have a bulk-boundary correspondence, where the ground state of the boundary changes fermion parity after absorbing a soliton.


Next, we must show that a loop with winding number 2 is contractible in $\CM_1$. One way to do it, from the perspective of the diabolical point, is to study two copies of $\eqref{AWmodelA1d}$, which we consider as four real fermions. One can find a $U(1)$-breaking mass term which completely gaps the theory. We will study this system in detail in Section \ref{subsecWZW1d}.

Another way to do it, from the perspective of the gapped region, is to study a different parametrization of \eqref{AWmodelA1d} over the circle $|M| = m > 0$, where our parameter winds around the origin twice. It does not make sense to extend such a parametrization to the disc $|M| \le m$. Indeed, the diabolical point at the origin determines the winding number, so any extension to a disc must have a different diabolical locus inside. Doubling the winding number like this may look like a trivial redefinition of the system, but we will show that after breaking $U(1)$ we can extend this family to a disc without closing the gap, {\it i.e.} without any diabolical locus.

To do so, let us regard the complex fermion as a pair of real (Majorana) fermions and add yet another massive Majorana fermion field. We can consider a more general family of free field theories describing three Majorana fermions:
\begin{equation}\label{Majorana1d}
\CL=-\psi^a i\slashed{\partial} \psi^a- i M_{ab}\psi^a_+\psi^b_-. 
\end{equation}
Here $\psi^a_+$ and $\psi^a_-$ are right-handed and left-handed components of $\psi^a$.
Hermiticity requires the mass matrix $M_{ab}$ to be real. If we regard $\psi^a_+$ and $\psi^a_-$ as triplets of $O(3)_R$ and $O(3)_L$, then the mass matrix transforms as $M\mapsto O_R^t MO_L$. One can use these transformations to make $M$ diagonal, with real positive entries. These entries determine the physical masses of excitations. The original one-parameter family of theories (\ref{AWmodelA1d}) is recovered when one of the physical masses becomes infinite. The rules of the game allow us to add a third Majorana fermion with a very large physical mass and then deform all the physical masses to be the same. Then the mass matrix has the general form $M=m O$, where $O\in O(3)$ and $m>0$. The space of orthogonal matrices has two components, each of which is topologically equivalent to $\RRR\PPP^3$. The original family is homotopic to a loop in one of these components. Since $\pi_1(\RRR\PPP^3)=\ZZZ_2$, any loop becomes contractible when iterated twice. Thus, our family over $S^1$ which winds twice may be extended to a family over a bounding disc, just using nonzero mass terms---that is, without closing the gap. This shows that loops with an even winding number become contractible when regarded as loops in $\CM_1$.\footnote{According to the classification in Section \ref{secgenremarks}, $\pi_1 (\CM_1) = \ZZZ_2$, so this pump invariant is again complete.}

This illustrates a general feature of topological families with torsion invariants (as opposed to ones like the $U(1)$ Thouless invariant, which is $\ZZZ$ valued). One may say that if the family looks nontrivial for some parameter $\theta \in [0,2\pi)$, but trivial if that family is defined for a $2\pi n$ periodic parameter, what is the physical difference? There is no physical difference of course, but to obtain the strongest conclusions from our framework one should use the ``minimal" parametrization which is relevant to the problem. This is similar to how one may see SPT phases in an spin-1 system like the Haldane chain by studying $SO(3)$ symmetry, which is natural if there are no half-integer spin degrees of freedom, but this SPT phase looks trivial if we consider the symmetry as $SU(2)$. For families presented by an action principle, and especially for families defined near a diabolical locus, there is always a minimal choice of periodicity.

We comment that the usual explanation of the robustness of the gapless point at $M=0$ involves a mixed 't Hooft anomaly between the vector and axial $U(1)$ symmetry (or, if the $U(1)$ symmetry is broken, an 't Hooft anomaly between fermion parity and axial $U(1)$). However, this assumes that the mass term is the only term which breaks axial $U(1)$. This assumption is not very natural, since axial $U(1)$ cannot be a microscopic on-site symmetry for any $M$ if we assume that at some UV scale the theory is described by a lattice Hamiltonian. Our argument does not rely on the axial $U(1)$ symmetry.

\subsection{The WZW term as a higher Berry phase and a codimension-4 diabolical point}\label{subsecWZW1d}

\subsubsection{A topologically nontrivial family}

Now we will discuss a family which is topological but which does not rely on any symmetry. In particular, there is no way to understand its nontriviality by any kind of pumping argument. However, many of the same features we explored for the Thouless pump in Section \ref{subseccomplexferms1d} will appear here. In many ways, this family is a 1+1d analogue of the classic example of the Berry phase of a spin-1/2 in a magnetic field.

The family we are interested in is defined by a pair of complex fermions $\psi^1, \psi^2$, packaged as a doublet $\Psi$, with the Lagrangian
\begin{equation}\label{AWmodelB1d}
\CL_\text{AW}=-\bar\Psi i\slashed{\partial} \Psi- i\bar\Psi(M^0+ i\gamma^{01}M^k\sigma^k)\Psi,
\end{equation}
where $\sigma^k$, $k = 1,2,3$ are Pauli matrices acting in the flavor space of $\Psi$. This model belongs the Abanov-Wiegmann ``B" series \eqref{eqnAWoddB}. There are four real parameters $M = (M^0,M^1,M^2,M^3)$. Any symmetries are accidental, and our conclusions will be robust to breaking them, but it is convenient to analyze this simple problem.

As before, this model is trivially gapped away from the massless point at the origin (it is a diabolical point). We study the response to slowly-varying parameters $\phi:X \to S^3_m$ on a sphere
\[S^3_m = \{M \in \RRR^4\ |\ |M| = m > 0\}.\]
Abanov and Wiegmann \cite{AbanovWiegmann} studied this problem and found that in the effective action for $\phi$, there is a Wess-Zumino-Witten term of level 1, which has the form
\begin{equation}\label{eqn1dberry}
S_\text{top}=\frac12 \int_X \eps^{\mu\nu} \omega^{(2)}_{ij}(\phi) \partial_\mu\phi^i\partial_\nu\phi^j d^2x=\int_X \phi^*\omega^{(2)},
\end{equation}
where $\omega^{(2)} = \frac12 \omega^{(2)}_{ij}(\phi) d\phi^i\wedge d\phi^j$ is locally a 2-form on $S^3_m$ satisfying $d\omega^{(2)} = 2 \pi {\rm vol}_{S^3_m}$ where ${\rm vol}_{S^3_m}$ is the homogeneous volume form on $S^3_m$ normalized to have volume 1.

More precisely, $\omega^{(2)}$ is a $U(1)$ 2-form gauge field on $S^3_m$ with Dixmier-Douady-Chern number 1, in other words with $2\pi$ units of magnetic flux through $S^3_m$. This means that from chart to chart, $\omega^{(2)}$ transforms by 1-form gauge transformations $\omega^{(2)} \mapsto \omega^{(2)} + d\lambda$, where $\lambda$ is a $U(1)$ gauge field. Note \eqref{eqn1dberry} is invariant under such transformations. A 2-form gauge background with Dixmier-Douady-Chern number $n$ on $S^3$ can be constructed by an analogue of the ``clutching construction" \cite{atiyah2018k}, where we define the 2-form on two hemispheres of $S^3$ and glue them by a gauge transformation along the equatorial $S^2$ by some $\lambda$ which has Chern number $n$ on the equator. This number may also be computed by integrating the curvature $\Omega^{(3)} = d\omega^{(2)}$ over $S^3_m$. We find it is
\begin{equation}\label{Hflux}
\Omega^{(3)}=\frac{1}{6\pi |M|^4}\eps^{ABCD} M^A dM^B dM^C dM^D.
\end{equation}
See Appendix \ref{appfreeferm} for a direct calculation. As a matter of fact, the model (\ref{AWmodelB1d}) has $Spin(4)$ symmetry with respect to which $\Psi_+$ and $\Psi_-$ transform as left-handed and right-handed spinors and $M^A$ transforms as a vector. This symmetry together with the equation $d\Omega^{(3)}=0$ fix the Berry curvature $\Omega^{(3)}$ up to a numerical factor.

The connection $\omega^{(2)}$ plays the role of the Berry connection for this system, in that it describes a quantized adiabatic response for this system (although it is second order in derivatives), so in this context we will refer to the Dixmier-Douady-Chern class of $\omega^{(2)}$, or equivalently the WZW level, as the higher Berry number. In fact it is directly related to the 0+1d Berry phase by dimensional reduction (see Appendix \ref{appdimred}). The coefficient of this term is quantized to be an integer by gauge invariance. We will give a direct argument for this in Section \ref{secbulkboundary} by relating it to the winding number of the phase of a certain family of partition functions.

As with the Thouless pump invariant, the presence of this quantized topological term protects the diabolical point inside $S^3_m$ from all perturbations. That is, there is no way to extend $\omega^{(2)}$ to a 4-ball filling $S^3_m$. We can therefore consider this family as representing a $\ZZZ \subset \pi_3 (\CM_1)$.\footnote{By the classification of Section \ref{secgenremarks}, this is an isomorphism.}

Because of the large symmetry of the model \eqref{AWmodelB1d}, the stability of the gapless point at $M=0$ can be explained using 't Hooft anomalies. The point $M=0$ has global $Spin(4)$ symmetry, and this symmetry cannot be gauged (it has an 't Hooft anomaly). This means that a gapless point at $M=0$ is robust with respect to $Spin(4)$-invariant perturbations.

The argument based on the higher Berry phase continues to apply even if we break this $Spin(4)$ symmetry. However, we cannot guarantee that the diabolical point does not turn into a different diabolical locus. For instance, if we add a chiral mass (not depending on the parameters $(\nu,n_i)$) to \eqref{AWmodelB1d}, then such a term will dominate near the origin, and gap the theory. However, there will still be some gapless points nearby. This situation is more generic in a 4-parameter phase diagram than the single gapless point of \eqref{AWmodelB1d}. The presence of gapless points is still protected by the higher Berry number, no matter what perturbations we add.

For a bosonic system also realizing this higher Berry phase, see Appendix \ref{appluttingerthouless}.

\subsubsection{Interfaces}

Let us examine the consequences of the higher Berry number for interfaces. Just as for the Thouless pump invariant, the higher Berry number implies that fixing a basepoint $M_* \in S^3_m$, any family of interfaces parametrized by $M \in S^3_m$, which connect the theory at $M_*$ for $x \to -\infty$ with the theory at $M$ for $x \to \infty$, must include at least one interface with a degenerate ground state. This follows from the same path space argument we used in Section \ref{subsubsec1dthoulessinterfaces}.

One simple family to analyze in detail is given by taking interfaces which interpolate along a straight line from $M$ to $M_*$. Then, the interface corresponding to $M = -M_*$ passes through the origin and corresponds to an interface hosting zero modes, while all other interfaces are trivially gapped\footnote{This can be seen perturbatively near the soluble interfaces $M = \pm M_*$. For the others this can be argued using the results of Appendix A of \cite{verresen2020topology}.}. Without loss of generality we can take $M_* = (m,0,0,0)$ and then this special interface corresponds to the inverting mass profile studied by Jackiw and Rebbi \cite{JackiwRebbi}. It has two zero modes localized to the place where the interface crosses the origin, therefore the ground-state has quadruple degeneracy.

Near this interface in the family, that is for $M = -M_* + \delta M$, we can see how the zero modes at the interface become gapped by studying perturbation theory. The effective Hamiltonian for the interface modes is
\begin{equation}
H_\text{eff}=a^\dagger_\alpha {\vec \sigma}^\alpha_\beta a^\beta \cdot {\vec \mu},
\end{equation}
where $a_\alpha,$ $\alpha=1,2$ are the annihilation operators for the interface modes, $\vec\sigma$ is the 3-vector of Pauli matrices, and $\vec \mu$ is defined by $\delta M = (0,\mu_1,\mu_2,\mu_3)$. This Hamiltonian has a non-degenerate ground state for all $\vec\mu\neq 0$ with energy $-2|\vec\mu|$. Moreover, from this 0+1d perspective, we find that the degenerate interface can be considered as a diabolical point!

This diabolical point, being codimension 3, should have some ordinary Berry curvature which protects it. To compute it, we note that both the Fock vacuum $\vert 0\rangle$ and the fully occupied state $a^\dagger_1a^\dagger_2\vert 0\rangle$ are eigenvectors of $H_\text{eff}$ with energy $0$ for all $\vec\mu$ and thus do not contribute to the Berry curvature. One may restrict  $H_\text{eff}$ to the two-dimensional subspace spanned by $a_\alpha^\dagger\vert 0\rangle$. In this subspace $H_\text{eff}$ reduces the the Hamiltonian of a spin-1/2 in a magnetic field $2\mu$. This is a paradigmatic example of a diabolical point, and the integral of the Berry curvature over the 2-sphere $|\vec\mu|=1$ is $2\pi$.

So we have seen that there is a single degenerate interface for $M = - M_* \in S^3_m$. Near this point, we can study an effective 0+1d problem and we find that this special interface represents a 0+1d diabolical point protected by the ordinary Berry curvature. However, this situation is not consistent from a 0+1d point of view, in that the small sphere which surrounds $M = -M_*$, which we have just argued has ordinary Berry number 1, is the boundary of a disc which closes off on the other side $S^3$, which would imply it has ordinary Berry number 0. In this sense, we consider the theory of the interface to have an anomaly: its peculiar phase diagram is only realizable because it exists as an interface. Following the terminology of Ref. \cite{Cordovaetal1}, the interface theory exhibits ``anomaly in the space of couplings".

We will see in Section \ref{secbulkboundary} that this is a general phenomenon that connects Berry phases and diabolical points in all dimensions. Moreover the number of diabolical points for the interface, weighted by their Berry numbers, is determined by the bulk higher Berry number, namely the integral of $\Omega^{(3)}$ over $S^3$.


The reason one gets quadruple degeneracy rather than the double degeneracy of von Neumann and Wigner is that our fermions are free. It is not possible to halve the number of fermions by imposing a Majorana condition on them. If we do this, the mass term in (\ref{AWmodelB1d}) will vanish. However, one can halve the degeneracy by adding an interaction of the form $\eps(\vec\mu) (a^\dagger a-1)^2$, where $\eps(0)>0$. It does not affect the subspace spanned by $a^\dagger_\alpha\vert 0\rangle$, but gives positive energy both to the Fock vacuum and the fully occupied state. 


\subsubsection{No Further Diabolical Points}\label{subsubsecnofurther}

Let us consider a system of $N$ real fermions in 1+1d with the Lagrangian (\ref{Majorana1d}). The mass matrix $M$ is real, and the theory is massive if $\det M$ is nonzero. Thus one can identify the parameter space of massive free theories as $GL(N,\RRR)$. It has two connected components labeled by the sign of the determinant of $M$ and denoted $GL_+(N,\RRR)$
and $GL_-(N,\RRR)$. They correspond to two different short-range entangled (SRE) phases of fermions in 1+1d, the nontrivial one being the Kitaev phase with Majorana zero modes at its boundaries. Each of the components is homotopy equivalent to $SO(N)$. For $N>2$ this space has fundamental group $\ZZZ_2$, which corresponds to our $\ZZZ_2$ fermion parity Thouless pump in Section \ref{subsubsecparitypump}. For $N > 4$, $\pi_3(SO(N)) = \ZZZ$ \cite{homotopygroups}, which corresponds to the codimension 3 diabolical point we have just discussed (see Appendix \ref{appfreeferm}).

There are further stable homotopy groups of $SO(N)$ in arbitrarily high degrees (exhibiting Bott periodicity), which would appear to imply there are free fermion diabolical points of arbitrarily high codimension. However, these points cannot be stable from the point of view of the effective action, since the WZW term \eqref{eqn1dberry} already has the most number of derivatives of the parameter background $\phi$ which can be fit in a topological term. This is also reflected in the classification of Section \ref{secgenremarks}. Thus, we expect all these free fermion families, and indeed any infrared-trivial family in $d$ space dimensions on $S^{m \ge d+3}$ can be extended to an infrared-trivial family on a bounding ball $B^{m+1}$, so there is no associated protected diabolical point. See Appendix \ref{appnofurther} for more details.

\section{Free fermions in 2+1d}\label{sec2+1d}

\subsection{Skyrmion charge and a codimension-3 diabolical point}\label{subsec2+1dthouless}

\subsubsection{A topologically nontrivial family}

\begin{figure}
    \centering
    \includegraphics[width=0.5\textwidth]{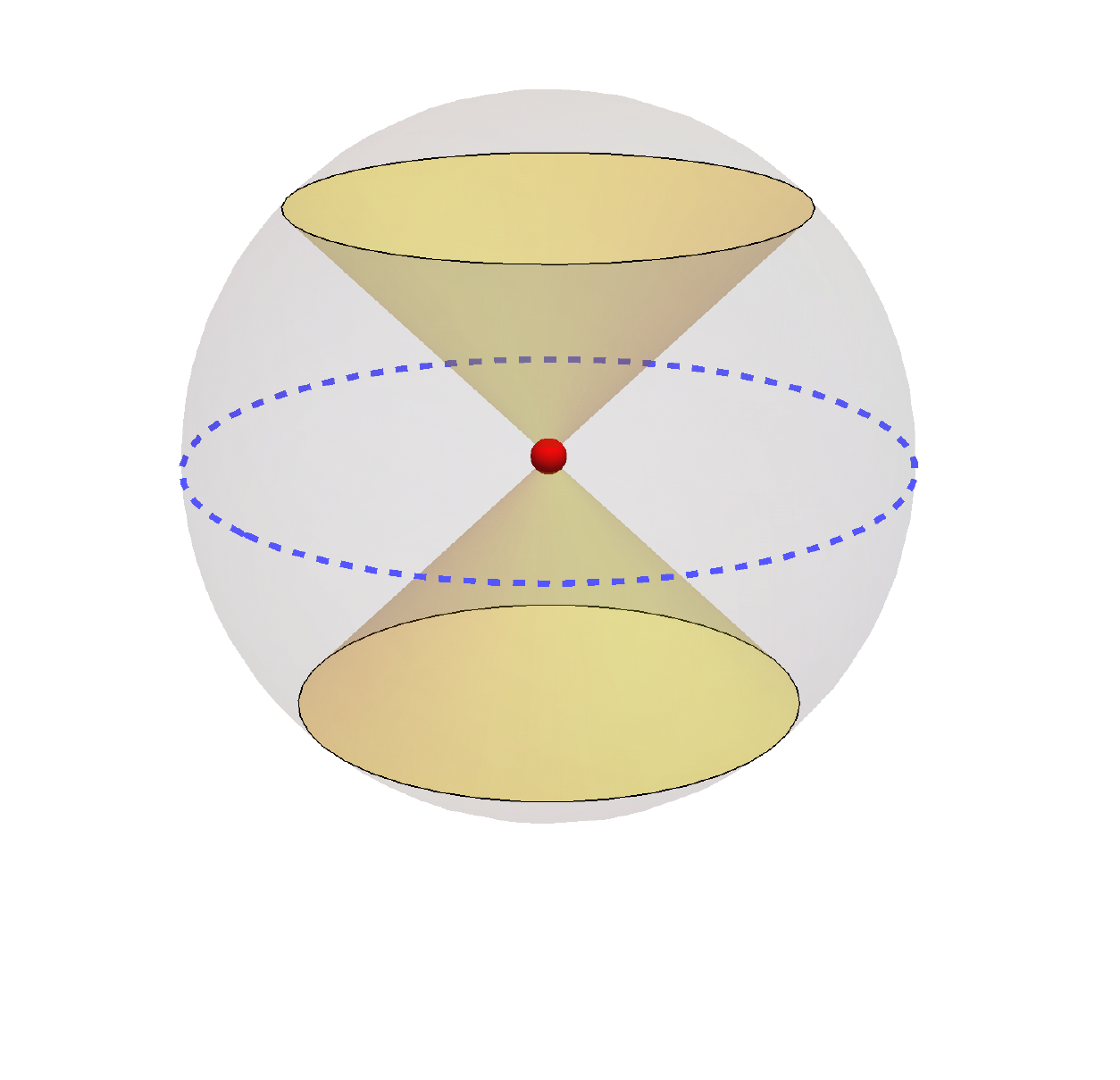}
    \caption{The nearby phase diagram of the model \eqref{AWmodelA2d}, with our two massless Dirac fermions at the origin (red dot), drawn with the vertical axis representing $\cos \nu$, and the horizontal axes schematically representing the three coordinates $n_i$. The dashed blue circle (actually an $S^2$) is the family at $\nu = \pi/2$ which realizes the 2+1d Thouless pump. The yellow cones are gapless points where one eigenvalue of the mass matrix changes sign. They separate the Thouless region from the contractible ``caps", where the system realizes the integer quantum Hall effect with Chern-Simons level ({\it i.e.} Chern number) $k = \pm 1$, depending on the sign of $\cos \nu$.}
    \label{figphasediagramtwodiracs2d}
\end{figure}

Now we consider two 2+1d Dirac fermions $\psi_{1,2}$ with $U(1)$ symmetry $\psi_j \mapsto e^{i\alpha}\psi_j$. We write the doublet $\Psi=(\psi_1,\psi_2)$.  The most general free Lagrangian which preserves $U(1)$ symmetry is 
\begin{equation}\label{AWmodelA2d}
\CL=-\bar\Psi i\slashed{\partial}\Psi-im\bar\Psi \left(\cos\nu+ n_i \sigma^i \sin\nu\right)\Psi
\end{equation}
where $\nu\in (0,\pi)$, $n^in^i=1$ and $\sigma^i$ are Pauli matrices acting in flavor space. $\nu$ and $n^i$ together parameterize $S^3$ with radius $m$ in $\mathbb{R}^4$. Without loss of generality we can take $m\geq 0$ by $n\rightarrow-n$, $\nu\rightarrow\pi-\nu$. The parameters $\sin \nu\, n^i$ for fixed $\nu$ can be thought of as living on an $S^2$ of radius $\sin \nu$. The model preserves $SU(2)$ symmetry at $\nu=0,\pi$ where this $S^2$ shrinks to zero size. We draw the phase diagram in Fig. \ref{figphasediagramtwodiracs2d}.

Consider first $\nu=\frac{\pi}{2}$. This family is the 2+1d sibling of \eqref{AWmodelA1d}, the next member of the Abanov-Wiegmann A series \eqref{eqnAWevenA}. For $m>0$ the theory is gapped with a unique ground state, while for $m=0$ the theory is gapless. 
One can ask if it is possible to remove the gapless point by a  $U(1)$-preserving deformation without bringing new gapless point from infinity. One may even allow deformations which involve additional massive degrees of freedom. We will now argue that this is not possible. If we promote $n$ to be position dependent and turn on the background gauge field $A$ for the $U(1)$ symmetry, then it can be shown \cite{AbanovWiegmann}  that 
the effective action contains\footnote{
In fact, one can use $SU(2)$ rotation $U(x)$ to fix $n$ to be a constant while introducing $SU(2)$ background gauge field $A=iU^{-1}dU$. Then the improperly quantized Chern-Simons term for the background $SU(2)$ gauge field reproduces the $\theta=\pi$ term.} a theta-term for $n$ with $\theta=\pi$ ({\it i.e.} the Hopf term) together with the following $A$-dependent term:
\begin{equation}\label{AWeff2d}
S_\text{top}=\frac{1}{8\pi}\int \eps^{\mu\nu\rho} A_\mu \eps^{ijk} n_i\partial_\nu n_j\partial_\rho n_k d^3 x.
\end{equation}
A derivation of the $A$-dependent term is given in Appendix \ref{appfreeferm}. The coefficients of these two terms are related by the spin-charge relation \cite{Freedhopf}.

Eq. (\ref{AWeff2d}) is a special case of the following effective action for a family of gapped  $U(1)$-invariant models with scalar parameters $\phi^i$:
\begin{equation}\label{Thouless2d}
S_\text{top}=\frac12 \int \eps^{\mu\nu\rho}A_\mu \partial_\nu\phi^i\partial_\rho\phi^j \tau_{2, ij}(\phi) d^2x=\int A\wedge\phi^*\tau_2,
\end{equation}
where $\tau_2=\frac12\tau_{2,ij}(\phi) d\phi^i\wedge d\phi^j$ is a 2-form on the parameter space. Compare \eqref{Thouless1d}. Such terms in the effective action were recently considered in the context of lattice models \cite{KapSpo2}. As described in \cite{KapSpo2}, they describe a 2+1d analog of the Thouless pump. 
Gauge-invariance requires $\tau_2$ to be closed and have integral periods. 
Equivalently, one can say that closedness and integrality is required by the conservation of the topological current and the integrality of the charge of skyrmions (topologically nontrivial configurations of the fields $\phi^i$ which approach a constant at infinity).

In the case of the model (\ref{AWmodelA2d}) at $\nu=\pi/2$, the form $\tau_2$ is proportional to the volume form of $S^2$, and the topological current takes the form 
\begin{equation}\label{eqn:3d2fcurrent}
J^\mu_\text{top}=\frac{1}{8\pi}\eps^{\mu\nu\rho}\epsilon^{ijk} n_i\partial_\nu n_j\partial_\rho n_k.
\end{equation}
The basic skyrmion is obtained by identifying $\RRR^2$ with a point at infinity added with $S^2$ and letting $n$ to be the identity map $S^2\ra S^2$. The charge of such a skyrmion is $1$.\footnote{In some settings this is referred to as quantum Hall ferromagnetism. See \cite{quantumhallferro} and references within.} 
Since the form $\tau_2$ is not exact, the 2-parameter family obtained by setting $\nu=\pi/2$ is not contractible in the space $\CMU_2$ of all infrared-trivial $U(1)$-invariant field theories in 2+1d.

The stability of the gapless point at $m=0$ can also be explained by the mixed 't Hooft anomaly between the time-reversal symmetry and the $SU(2)$ flavor symmetry which acts on $\Psi_1,\Psi_2$ as a doublet. Time-reversal symmetry is present for all $n$ if $\nu$ is fixed to be $\pi/2$, while $SU(2)$ symmetry is restored at $m=0$. Therefore 't Hooft anomaly matching requires gapless modes at $m=0$. However, this explanation only shows stability with respect to deformations which preserve both symmetries. For example, this kind of argument breaks down if $\nu$ is slightly different from $\pi/2$. Also, the mixed 't Hooft anomaly between time-reversal symmetry and $SU(2)$ has order $2$, {\it i.e.} it becomes trivial if we take an even number of copies of the system (\ref{AWmodelA2d}) with $\nu=\pi/2$. The argument based on the effective action (\ref{Thouless2d}) shows that neither the number of copies, nor time-reversal nor flavor symmetry are essential for the stabilty of the gapless point at $m=0$.

\subsubsection{Interfaces}

This non-contractibility has immediate consequences for smooth interfaces between different member of the family. We can fix $n=(0,0,1)$ as the basepoint and consider smooth interfaces between the basepoint model and other members of the $\nu=\pi/2$ family obtained by varying $mn^i$ along a straight line segment. By a general argument explained in Section \ref{secbulkboundary}, if this family of interfaces is continuous, at least one of the interfaces must be gapless. For the model (\ref{AWmodelA2d}) a gapless interface occurs for $n=(0,0,-1)$. This is explained in detail in Appendix \ref{appdimred}. In brief, one finds that the eigenvalue equation for modes with zero momentum along the interface reduces to the 1d eigenvalue equation for the Jackiw-Rebbi model ({\it i.e.} the model (\ref{AWmodelA1d}) with $M_2=0$ and $M_1$ varying between $m$ and $-m$). Thus there is exactly one zero mode in this case, and the interface carries a massless Dirac fermion.

Now consider perturbing $n$ from $(0,0,-1)$ to a nearby point $\left(w_1,w_2,-1+O(w_{1,2}^2)\right)$. Using perturbation theory in $w_1,w_2$, it is easy to check that this gaps the Dirac fermions. The low-energy effective action for the interface mode is then described by Eq. (\ref{AWmodelA1d}) with $M_1=w_1, M_2=w_2$. After integrating out the mode on the interface, we end up with an effective action (\ref{Thouless1d}), where the 1-form $\tau_1$ on the $w$-plane is given by
\begin{equation}
\tau_1=\frac{1}{2\pi}d\, {\rm arg} w,
\end{equation}
where $w=w_1+iw_2$.
This form is singular at $w=0$ which is the location of the interface diabolical point. Thus a codimension-3 diabolical point for the bulk theory leads to a codimension-2 diabolical point for the interface, which is again protected by a Thouless pump.

There is also a point $n=(0,0,1)$ where there are no interface modes at all. Note that the problem studied here preserves a $U(1)$ subgroup of $SU(2)$ flavor symmetry, and that $n=(0,0,1)$ and $n=(0,0,-1)$ are the only two points on $S^2$ fixed by this symmetry. Thus, if we assume that the gapless modes on the interface occurs only at isolated points in the parameter space (which is a generic situation), then $n=(0,0,-1)$ is the only such point. 

Note that if we had a UV-complete theory describing the interface for all values of $n$, then the sum of periods of $\tau_1$ over all gapless points would necessarily be zero. Here it is equal to $1$, thanks to the existence of a point where the interface mode becomes non-normalizable and escapes to the bulk. In the language of Ref. \cite{Cordovaetal1}, the interface theory has an ``anomaly in the space of couplings". 

\subsubsection{A more general family}

Now let us move away from $\nu=\frac{\pi}{2}$. Since $S^3$ does not have non-contractible 2-cycles, we conclude there must be a phase transition as we move away from $\nu=\frac{\pi}{2}$ while keeping the sphere radius $m$ fixed. Indeed, since the determinant of the mass matrix is $m^2\cos(2\nu)$, there are two cones of gapless field theories at $\nu=\frac{\pi}{4}$ and $\nu=\frac{3\pi}{4}$. They meet at the codimension-4 diabolical point $m=0$. See Fig. \ref{figphasediagramtwodiracs2d}.


Let us briefly comment on what protects the codimension one gapless cones at $\nu=\frac{\pi}{4},\nu=\frac{3\pi}{4}$. These are also diabolical points, in a sense, although they are more familiar as a continuous transition between different integer quantum Hall states.
To see this, note the masses of the two Dirac fermions are given by the eigenvalues of the mass matrix, which are
\begin{equation}
\sqrt{2} m\cos\left(\nu \pm \frac{\pi}{4}\right)~.
\end{equation}
Thus, due to the parity anomaly, whenever the mass of one of the Dirac fermion crosses from negative to positive the effective action contains additional term $\frac{1}{4\pi}AdA+2\text{CS}_\text{grav}$ where $\int_{M_3}\text{CS}_\text{grav}=\pi\int_{M_4}\hat A(R)$ with $M_3=\partial M_4$ is the gravitational Chern-Simons term.

If we study an interface which varies from $\nu < \pi/4$ to $\pi/4 < \nu < 3\pi/4$, these Chern-Simons terms lead to gapless chiral modes near where $\nu$ crosses $\pi/4$. This can be thought of as another manifestation of bulk-boundary correspondence.

In light of this, the Abanov-Wiegmann theory does not describe a generic codimension 3 diabolical locus. Suppose we begin with the 3-parameter phase diagram of \eqref{AWmodelA2d} with $\nu = \pi/2$, just parametrized by the $n_i$, and then tune $\nu$ slightly away from $\pi/2$. Because the diabolical point at the origin is protected by the Thouless invariant, it cannot disappear completely, but it is unstable in the view of the above. For small $n_i$, the $SU(2)$-invariant mass term dominates, and gaps the central point. We find that it is replaced by an island of an integer quantum Hall state with a single gapless Dirac fermion along its two-dimensional boundary. For a 3-parameter phase diagram, these diabolical islands are the generic situation, provided $U(1)$ symmetry is maintained.

\subsection{The Wess-Zumino-Witten term and a codimension-5 diabolical point}

Even if we do not assume any symmetry, a possible topological term in the effective action for scalar fields $\phi^i$ is the WZW term
\begin{equation}\label{WZW2d}
S_\text{top}=\int_X \phi^* \omega^{(3)},
\end{equation}
where $\omega^{(3)}$ is a $U(1)$ 3-form gauge field. Compare \eqref{eqn1dberry}. Locally it is a 3-form, but it undergoes a 2-form gauge transformation $\omega^{(3)}\mapsto \omega^{(3)}+dB$ when one changes charts, under which \eqref{WZW2d} is invariant. Here $B$ is a 2-form gauge field, similar to $\omega^{(2)}$ in Section \ref{subsecWZW1d}. Therefore the 4-form $\Omega^{(4)}=d\omega^{(3)}$ is well-defined and closed, but not necessarily exact. Moreover, for $\exp(iS_\text{top})$ to be well-defined, periods of $G$ must satisfy certain quantization conditions. For fermionic systems with a general parameter space these conditions are rather nontrivial, see Ref. \cite{KapSpo1} for a brief  discussion. However in the case when the parameter space is $S^4$ the same argument as for the 1+1d WZW term shows that $\int_{S^4}\Omega^{(4)}$ must be an integral multiple of $2\pi$. If the cohomology class of $\Omega^{(4)}$ is nontrivial, the corresponding family of field theories in non-contractible in $\CM_3$, the space of infrared-trivial 2+1d field theories.

As a simple example, consider a theory four complex fermions $\Psi=(\psi_1,\psi_2,\psi_3,\psi_4)$. It is convenient to regard them as a spinor representation of a flavor $Spin(5)$ symmetry, where $\Psi$ transforms as a spinor, even though this symmetry will not be important in our analysis. Following Ref. \cite{AbanovWiegmann} we consider the Lagrangian
\begin{equation}\label{AWmodelB2d}
\CL=-\bar\Psi i\slashed{\partial}\Psi    -i m\bar\Psi\left(\cos\nu\Gamma^5+n_i\Gamma^i\sin\nu  \right)\Psi~,
\end{equation}
where $\Gamma^i$ are the gamma matrices for the $Spin(5)$ flavor indices, and $n^2=1$, $i=1,2,3,4$.
Under the $Spin(5)$ flavor action on $\Psi$ the mass parameters $(m\cos\nu,m\sin\nu\, n^i)$ transform as a vector. Without loss of generality we take $m\geq 0$ by $\nu\rightarrow \pi-\nu,n\rightarrow -n$. There is also a $U(1)$ symmetry under which $\Psi\mapsto e^{i\gamma}\Psi$.

For $m>0$ the theory is gapped with a unique ground state. The point $m=0$ is a gapless diabolical point of codimension 5. The parameter space with this point removed is homotopy equivalent to $S^4$. This diabolical point is protected by the higher Berry phase on $S^4$ and thus is stable under arbitrary deformations, including those which break $Spin(5)$ and/or $U(1)$. 
Indeed, if we promote the parameters $\phi=(\nu,n_i)$ to be position dependent, the effective action of the fermion contains \cite{AbanovWiegmann} (see also Appendix \ref{appfreeferm}) the WZW term (\ref{WZW2d}) with  
\begin{equation}
\Omega^{(4)}=\frac{1}{8\pi}\epsilon^{ijkl}\sin^3\nu d\nu n_i dn_j dn_k dn_l~.
\end{equation}
This is the volume form of $S^4$ normalized so that $\int_{S^4} \Omega^{(4)} =2\pi$.
Since this integral is quantized, the family (\ref{AWmodelB2d}) is not contractible in the space of infrared-trivial 2+1d field theories.

As in Section \ref{subsec2+1dthouless}, if we add a constant diagonal mass term to \eqref{AWmodelB2d}, this will open a gap near $M = 0$ into a topological phase (equivalent to 8 copies of $p+ip$ superconductors). At the boundary of this region two of the eigenvalues of the mass matrix change sign, yielding an $S^4$'s worth of two massless Dirac fermions, which is protected by the higher Berry number of the surrounding region.

A discussion of interfaces in this model and other models in the ``B'' series of Abanov and Wiegmann can be found in Appendix \ref{appdimred}.


\section{Free fermions in 3+1d}\label{sec3+1d}

\subsection{Skyrmion charge and a codimension-4 diabolical point}

Consider a model of free fermions in 3+1d with a Lagrangian
\begin{equation}\label{Dirac3d}
\CL=-\bar\Psi i\slashed{\partial}\Psi-i \bar\Psi (M^0+i\gamma^{0123} M^i\sigma^i)\Psi,
\end{equation}
where $\Psi=(\psi^1,\psi^2)$ is a doublet of Dirac fermions, $\gamma^{0123}=i\gamma^0\gamma^1\gamma^2\gamma^3$ is the chirality operator, $\sigma^i$ for $i=1,2,3$ are Pauli matrices, and $M=(M^0,M^1,M^2,M^3)$ is the 4-vector of mass parameters. Uppercase Roman indices take values from $0$ to $3$. This is a 3+1d analog of the model (\ref{AWmodelA1d}). Its topological properties  have been studied by Goldstone and Wilczek \cite{GoldstoneWilczek}. It is also a special case of the ``A'' series of Abanov and Wiegmann \cite{AbanovWiegmann}. The model is gapless for $M=0$ and gapped otherwise. 

We claim that $M=0$ is a codimension-4 diabolical point protected by $U(1)$ symmetry. To see this, we need to examine topological terms in the effective action which depend on the background $U(1)$ gauge field $A$ as well as scalar parameters $\phi^i$ representing the slowly-varying mass terms $M$. The term which is analogous to (\ref{Thouless1d}) and (\ref{Thouless2d}) is linear in $A$ and has the form
\begin{equation}\label{Thouless3d}
\begin{gathered}
S_\text{top}=\frac{1}{6}\int_X \eps^{\mu\nu\rho\sigma} A_\mu \partial_\nu\phi^i  \partial_\rho\phi^j \partial_\sigma\phi^k \tau_{3,ijk}(\phi) d^4x \\ =\int_X A\wedge\phi^*\tau_3,
\end{gathered}
\end{equation}
where $\tau_3$ is a 3-form on the parameter space. Compare \eqref{Thouless1d} and \eqref{Thouless2d}. Gauge invariance requires $\tau_3$ to be closed and to have integral periods. If the restriction of $\tau_3$
to some 3-parameter family is not exact, the corresponding family cannot be contracted to a point in the space $\CMU_3$ of infrared-trivial 3+1d field theories with a $U(1)$ symmetry. Intuitively, this is because the term (\ref{Thouless3d}) gives rise to a topological current
\begin{equation}
J^\mu_\text{top}=\frac{1}{6}\eps^{\mu\nu\rho\sigma}\partial_\nu\phi^i\partial_\rho\phi^j\partial_\sigma\phi^k \tau_{3,ijk}(\phi)
\end{equation}
which gives charge to skyrmions (topologically nontrivial configurations of the parameter fields $\phi^i(x)$). Continuously deforming the model within the class of infrared-trivial field theories cannot change the charge of skyrmions. Note that charge quantization for arbitrary spatial slices is equivalent to the integrality of periods of $\tau_3$.

In the case of interest to us, the parameters $\phi^i$ can be identified with $M^A$. The gapped locus is defined by $M\neq 0$ and is homotopically equivalent to $S^3$. This space has $H^3(S^3,\ZZZ)= \ZZZ$, so there is a possibility for a term of the form (\ref{Thouless3d}). 
In fact, it was shown in \cite{GoldstoneWilczek} that upon integrating out fermions the model (\ref{Dirac3d}) generates the effective action (\ref{Thouless3d}) with the 3-form $\tau_3$ given by
\begin{equation}
\tau_3=\frac{1}{12\pi^2|M|^4}\eps^{ABCD} M^A dM^B dM^C dM^D.
\end{equation}
This implies that the family (\ref{Dirac3d}) describes a codimension-4 diabolical point. 

The consequences of this for interfaces are similar to the ones in the 1+1d and 2+1d cases. Namely, if we restrict to the $S^3$ given by $|M|=1$ and study interfaces from a fixed basepoint $p_0$ to other points $p \in S^3$, then any family of interfaces which depends continuously on $p$ will have at least one gapless interface. If there is a single such point $p_* \in S^3$, we can consider it as a 2+1d diabolical point for the interface theory, and the effective action for the nearby gapped interfaces will contain a topological term of the form \eqref{Thouless2d}, where the 2-form $\tau_2$ integrates to $1$ on any $S^2$ surrounding $p_*$. That is, the field theory on the interface has an ``anomaly in the space of couplings" in the sense of Ref. \cite{Cordovaetal1}. For a certain nice family of interfaces, the effective 2+1d theory on the interface can be described by the Lagrangian \eqref{AWmodelA2d}. This is explained in more detail in Appendix \ref{appdimred}.

\subsection{WZW terms and a codimension-6 diabolical point}

If we do not assume any symmetry, then the most obvious topological term in the effective action is the WZW term
\begin{equation}\label{WZW3d}
S_\text{top}=\int_X \phi^* \omega^{(4)},
\end{equation}
where $\omega^{(4)}$ is a 4-form gauge field on the parameter space, analogous to \eqref{eqn1dberry}, \eqref{WZW2d}, and the quantum mechanical Berry phase. That is, locally $\omega^{(4)}$ is a 4-form, but when one goes from chart to chart $\omega^{(4)}$ transforms by a 3-form gauge transformation $\omega^{(4)}\mapsto\omega^{(4)}+dC$, where $C$ is a 3-form gauge field. \eqref{WZW3d} is invariant under such gauge transformations. The gauge curvature $\Omega^{(5)}=d\omega^{(4)}$ is a closed but not necessarily exact 5-form on the parameter space (the WZW 5-form). 


As in lower dimensions, if $\Omega^{(5)}$ has a nonzero integral over a family parametrized by $S^5$, then that family cannot be contracted to a point within the space $\CM_3$ of infrared-trivial field theories in 3+1 dimensions. In particular, we expect that there are diabolical points in codimension 6 which are stable without imposing any symmetry. An example of such a model has been constructed by Abanov and Wiegmann \cite{AbanovWiegmann}. Consider four Dirac fermions $\Psi=(\psi^1,\ldots,\psi^4)$ transforming in the fundamental representation of $USp(4)$ (or equivalently, as a spinor of $Spin(5)$). Following Ref. \cite{AbanovWiegmann} we consider the Lagrangian
\begin{equation}\label{AWmodelB3d}
\CL=-\bar\Psi i\slashed{\partial}\Psi-i \bar\Psi (M^0+i\gamma^{0123} M^i \Gamma^i)\Psi,
\end{equation}
where $\Gamma^i$, $i=1,\ldots,5$, are $Spin(5)$ Dirac matrices acting in the flavor indices and $M=(M^0,M^1,\ldots,M^5)$ is a 6-vector of mass parameters. The model has a gapless point at $M=0$. Ref. \cite{AbanovWiegmann} shows that upon integrating out the fermions one gets the effective action  (\ref{WZW3d}) where the 5-form $\Omega^{(5)}$ on $\RRR^6\backslash \{0\}$ is given by
\begin{equation}
\Omega=\frac{1}{60\pi^2|M|^6}\eps^{ABCDEF} M^A dM^B dM^C dM^D dM^E dM^F
\end{equation}
The integral of this 5-form over the $S^5$ is $2\pi$. This implies that $\Omega^{(5)}$ is not exact and therefore the family (\ref{AWmodelB3d}) is not contractible in the space of all 3+1d infrared-trivial field theories.

As usual, non-contractibility implies that a family of interfaces which depends continuously on $M$ will have at least one gapless point. As explained in Appendix C, in the case of model (\ref{AWmodelB3d}) such a gapless point occurs when one considers an interface between $M=(m,0,0,0,0,0)$ and $M=(-m,0,0,0,0,0)$. In the neighborhood of this special interface there are almost-gapless interface modes described by the Lagrangian (\ref{AWmodelB2d}). This is an example of the boundary-bulk correspondence.

\subsection{Axion couplings}\label{subsecaxions}

Until now, all diabolical loci not protected by any symmetry arose from WZW terms in the effective action for the parameter fields. Such diabolical loci occur in codimension $d+3$, where $d$ is the spatial dimension, and are direct descendants of the von Neumann-Wigner points. In 3+1d a new phenomenon occurs: one encounters diabolical loci in codimension $2$ which are related to coupling to background geometry. Topological invariance requires these terms to have the form
\begin{equation}
S_\text{top}=\frac{1}{2\pi}\int_X \text{CS}_\text{grav}\wedge \phi^*\rho,
\end{equation}
where $\rho$ is a 1-form on the parameter space with quantized periods. One can always assume that the parameter space is $S^1$, in which case $\rho= N d\alpha$, where $\alpha$ is a $2\pi$-periodic coordinate on $S^1$ and $N$ is a number. Then the above coupling can be written as an axion coupling
\begin{equation}\label{axion}
S_\text{top}=-\frac{N}{384\pi^2}\int_X \alpha\, {\rm Tr} R^2.
\end{equation}
In order for $\exp(i S_\text{top})$ to be well-defined, $N$ has to be quantized. The precise quantization condition depends on whether $X$ is allowed to be an arbitrary oriented 4-manifold (the bosonic case) or a spin manifold (the fermionic case). In the former case, $N$ must be a multiple of 16, while in the latter case it must be integral. Whenever $N$ is nonzero, the family parameterized by $\alpha$ is non-contractible, signifying the presence of a diabolical locus in codimension $2$. This diabolical locus is stable with respect to arbitrary modifications of the theory in the UV. 

Examples of models where gravitational axion couplings arise are well-known. One can take a single Majorana fermion with mass $M$ and set $\alpha={\arg\, M}$. Gravitational anomaly for chiral $U(1)$ symmetry then leads to the coupling (\ref{axion}) with $N=1$. 

The physical consequences of the gravitational axion coupling are also  well-known. If $\alpha$ winds $k$ times around the origin in the $12$ coordinate plane in spacetime, then at the origin of the coordinate plane there must be chiral gapless modes propagating in the $x^3$ direction with the chiral central charge $c_R-c_L=Nk/2$. If one considers smooth interfaces between all models with $|M|=m>0$ and the basepoint model $M=m$, then there will be a gapless interface for at least one value of $\alpha={\rm arg}\, M$. This is so even if one modifies the theory in the UV while preserving  the gap for all $\alpha$. In the case of the massive 3+1d Majorana fermion, the gapless interface carries a massless 2+1d Majorana fermion. 

Usually the stability of the gapless point at $M=0$ is explained in terms of the 't Hooft anomaly for the chiral $U(1)$ symmetry. This symmetry is restored at $M=0$ and thus 't Hooft anomaly matching requires the presence of massless modes. This explanation is problematic from the point of view of lattice regularization, since a chiral $U(1)$ symmetry cannot be realized on a lattice in a completely local way. Our explanation avoids using any symmetries and relies instead on the topology of the parameter space of infrared-trivial field theories. 

If one considers gapped models with a global $U(1)$ symmetry, an axion coupling to a background $U(1)$ gauge field is also allowed. Such a coupling signifies a diabolical locus in codimension $2$ as well. The usual explanation of stability involves a mixed 't Hooft anomaly between two different $U(1)$ symmetries one of which is broken only by a mass term.

\section{Interacting Gauge Theories}\label{secgaugetheories}

\subsection{Gauge theories in 2+1d}

\subsubsection{Operation \texorpdfstring{$ST$}{ST} and the generalized Thouless Pump Invariant}


Let us describe a general technique for producing new diabolical loci in 2+1d protected by a $U(1)$ Thouless pump. Suppose we begin with such a family. If we turn on nontrivial background gauge field $A$ for the $U(1)$ symmetry and promote the parameters to be position-dependent $\phi:X\rightarrow {\cal M}$ on spacetime $X$, then the effective action can be expressed as (\ref{Thouless2d}). 
The $U(1)$ current is 
\begin{equation}\label{eqn:tau2current}
j^\mu=\frac{1}{2}\epsilon^{\mu\nu\rho}\partial_\nu\phi^i\partial_\rho\phi^j\tau_{2,ij}(\phi)~.
\end{equation}


Now, let us gauge the $U(1)$ symmetry and study the effective action of the resulting theory.
We first add a Chern-Simons counterterm
\begin{equation}
\int_X\left(\frac{1}{4\pi}AdA-\frac{1}{2\pi}AdB\right)
\end{equation}
where $B$ is a background $U(1)$ gauge field, and then promote $A$ to be a dynamical gauge field $a$. We may also add a Maxwell term for $a$, but the infrared behavior is determined by the level one Chern-Simons term for $a$. It makes $a$ massive, and more over makes the corresponding low-energy theory invertible (see for instance Appendix B of \cite{Seiberg:2016gmd}).
Thus we obtain a new system that is invertible away from the diabolical points and has a global $U(1)$ symmetry with a current
\begin{equation}\label{eqn:topcurrent}
j'^\mu=\frac{1}{2\pi}\epsilon^{\mu\nu\rho}\partial_\nu a_\rho
\end{equation}
which couples to the background gauge field $B$.
This is the $ST$ operation of Ref.~\cite{Witten:2003ya} which originally appeared in \cite{Kapustin:1999ha}.

The new system is described at low energies by a topological $U(1)$ gauge theory with an action
\begin{equation}\label{eqn:3deffa}
S_\text{eff}=\int_X\left(\frac{1}{4\pi}ada-\frac{1}{2\pi}adB+ a\, \phi^*\tau_2~\right).
\end{equation}
By shifting $a\mapsto a+B$ in (\ref{eqn:3deffa}) we find the effective action
\begin{equation}\label{eqn:seffnew}
S'_\text{eff}
=\int_X \left(\frac{1}{4\pi} ada+a\,\phi^*\tau_2\right)+\int_X \left(-\frac{1}{4\pi} BdB+B\,\phi^*\tau_2\right).
\end{equation}
The first term describes a family of infrared-trivial 2+1d field theories which does not couple to the background $U(1)$ gauge field $B$.\footnote{
The partition function of this invertible theory is a Hopf-like term
\begin{equation}
e^{iH[\phi]}\equiv 
\int Da e^{i\int_X \left(\frac{1}{4\pi} ada+a\,\phi^*\tau_2\right)}~,
\end{equation}
which can be formally expressed as (omitting the gravitational Chern-Simons term $-\int_X 2\text{CS}_\text{grav}$) $H[\phi]=\pi\int_Y\phi^*(\tau_2)^2$ for spin 4-manifold $Y$ that bounds the spacetime. It does not depend on the choice of $Y$. An example of such a Hopf term is discussed in \cite{Wilczek:1983cy}.}
Therefore the Thouless pump invariant is determined by the second term and is the same as for the original family 
(\ref{Thouless2d}), 
but diabolical point is modified: it is mapped to its $ST$ transform.


\subsubsection{\texorpdfstring{$U(1)_1$}{U(1)_1} with two fermions}

Let us apply the above procedure to the theory of two complex 2+1d fermions with $SU(2)$-covariant vector mass discussed in section \ref{sec2+1d}. In this case $j$ in (\ref{eqn:tau2current}) is the current for the skyrmion number.
The $U(1)$ symmetry transforms the two fermions by the same phase.

The $ST$ operation gives an interacting $U(1)_1$ Chern-Simons-matter theory with two fermions of charge one. This theory has a magnetic $U(1)$ symmetry whose current is the dual of the gauge field strength. 
From the previous discussion, away from the $m=0$ point the theory has a Thouless pump invariant for this $U(1)$ symmetry. 

The first term in the effective action (\ref{eqn:seffnew}) gives an additional $\theta=\pi$ Hopf term  \cite{Wilczek:1983cy} which cancels the Hopf term from two complex fermions discussed in section \ref{sec2+1d}.
This implies that the skyrmion that carries a unit charge of the topological current $j$ is a boson \cite{Wilczek:1983cy}.
Since the equation of motion for $a$ in (\ref{eqn:seffnew}) identifies the current (\ref{eqn:tau2current}) with (\ref{eqn:topcurrent}) up to a contact term, the skyrmions correspond to the monopoles in the gauge theory.
Indeed, the monopole operators in the $U(1)$ gauge theory with charge one fermions are bosons  \cite{Borokhov:2002ib}.\footnote{
Another way to see this is that the theory obeys spin/charge relation with respect to the dynamical $U(1)$ gauge field, and thus gauge invariant local operators are bosons.
} The critical point at $m=0$ is protected by the nonzero skyrmion charge as in the original family of two complex fermions.

Similar to the discussion in Section \ref{sec2+1d}, we can take $m>0$ and consider a family of interfaces with fixed base point $n=(0,0,1)$ obtained from the family in Section \ref{sec2+1d} by gauging the $U(1)$ symmetry.
 From the ``boundary-bulk'' correspondence this family must contain an gapless interface. An analysis along the lines of Appendix \ref{appdimred} shows that this gapless  interface hosts a massless periodic scalar which is the bosonization of a massless Dirac fermion.

\subsubsection{\texorpdfstring{$U(N)$}{U(N)} gauge theory with two scalars, deconfined quantum criticality, and boson/fermion duality}\label{sec:scalarThouless}

In the examples we discussed so far (and the examples discussed in \cite{AbanovWiegmann}) the Berry phase or Thouless pump invariant are derived from integrating out massive fermions. Here we present a new class of examples where the Berry phase is obtained from a massive scalar through the Higgs mechanism.

Consider $U(1)$ gauge theory with two complex scalars $\phi$ of charge one and quartic potential that respects the $SU(2)$ flavor symmetry\footnote{The more precise global symmetry is discussed in \cite{Benini:2017dus}}, aka the $\mathbb{CP}^1$ or abelian Higgs model.
We can add the $SU(2)$ vector mass term
\begin{equation}\label{eqn:scalarpotential}
V(\phi)=m^2 n_i\phi^\dag\sigma^i\phi+\lambda (\phi^\dag\phi)^2~,
\end{equation}
where we take $m^2\rightarrow \infty$, $\lambda>0$ and $\sigma^i$ are the Pauli matrices for $SU(2)$ isospin and $\sum n_i^2=1$.

For large $m^2$, due to the Higgs potential the $U(1)$ gauge field obtains a mass square $m^2/\lambda$. If we promote the mass parameters $n_i$ to be position dependent, then to leading order in the $1/m$ expansion there is current
\begin{equation}
j=i\left(\phi^\dag d \phi-\text{h.c.}\right)=-\frac{n_1d n_2-n_2 d n_1}{2(1-n_3)(\lambda/m^2)}+\cdots~,
\end{equation}
where $\cdots$ are suppressed by $1/m^2$.
The equation of motion for the $U(1)$ gauge field $a$ to leading order gives
\begin{equation}
a = \frac{n_1 d  n_2-n_2 dn_1} {2(1-n_3)}+\cdots~,
\end{equation}
where $\cdots$ are suppressed by $1/m^2$.
Thus to leading order,\footnote{
A useful identity is
\begin{equation}
n_3\left(n_1dn_2dn_3+n_2dn_3dn_1+n_3dn_1dn_2\right)
=dn_1dn_2~,
\end{equation}
which can be derived using $n_1^2+n_2^2+n_3^2=1$ and $n_3dn_3=-n_1dn_1-n_2dn_2$.
}
\begin{align}\label{eqn:uone2b}
da =&\frac{dn_1 dn_2}{2(1-n_3)}+\left(
\frac{dn_1 dn_2}{2(1-n_3)}+
\frac{dn_3(n_1 d n_2-n_2 dn_1)}{2(1-n_3)^2}
\right)\cr
=&-\frac{1}{2}\epsilon^{ijk}n_idn_jdn_k~,
\end{align}
The relation (\ref{eqn:uone2b}) implies that the monopole of the microscopic gauge theory corresponds to the skyrmion configuration of the field $n_i$.

If we turn on background gauge field $A$ for the magnetic $U(1)$ symmetry with current $j'^\mu=-\frac{1}{ 2\pi}\epsilon^{\mu\nu\rho}\partial_\nu a_\rho$, at low energy the theory contains the following effective action
\begin{equation}
\int A_\mu  j'^\mu=\frac{1}{8\pi}\epsilon^{\mu\nu\rho}\epsilon^{ijk}\int A_\mu n_i\partial_\nu n_j\partial_\rho n_k~.
\end{equation}
Thus this model provides an example of family of interacting bosonic theories that has a nontrivial 2+1d Thouless pump invariant.

For $m^2=0$ the model is proposed to describe the deconfined quantum critical point between the N\'eel phase and the valence bond solid (VBS) \cite{Senthil:2010ogy}. This is a proposed continuous phase transition in a $S = 1/2$ anti-ferromagnet from a N\'eel phase, where spin rotation $SO(3)$ is spontaneously broken, to a VBS phase, where site-rotation symmetry is broken. On a square lattice, this group is $\ZZZ_4$, on a honeycomb lattice, it is $\ZZZ_3$, and so on. The $SO(3)$ spin-rotation is proposed to act as the flavor symmetry of the two scalars, while rotations act as a subgroup of the $U(1)$ monopole symmetry.

The single symmetric relevant operator which tunes the transition is the $SO(3)$-invariant mass term $M^2\phi^\dag\phi$ for the scalars. For $M^2 \gg 0$, the scalars may be integrated out and we are left with a  free $U(1)$ gauge theory. If there are no monopoles in the action, this leads to a Coulomb phase where the $U(1)$ monopole symmetry is spontaneously broken. However, in the setting of Neel-VBS, only a finite subgroup, {\it e.g.} $\ZZZ_4$ or $\ZZZ_3$, is a microscopic symmetry, so generically there will be charge 4 or 3 (resp.) monopoles in the action, which are irrelevant at the critical point but destabilize the Coulomb phase, leading to a gapped phase where the rotation symmetry is spontaneously broken---this is the VBS phase.

For $M^2 \ll 0$, the scalars condense, and the $U(1)$ gauge field is Higgsed, with a vacuum moduli space $\mathbb{CP}^1 = S^2$, which means the $SO(3)$ symmetry is spontaneously broken and the system is in the N\'eel phase. This direct order-to-order transition may be explained for square, rectangular, or triangular lattices, which have a $\ZZZ_2$ site-rotation symmetry, by the 't Hooft anomaly of $SO(3) \times U(1)$ \cite{Benini:2017dus,Komargodski_2019,Metlitski_2018}. This anomaly argument is a version of the Lieb-Schultz-Mattis (LSM) theorem. Whether there is actually a continuous phase transition remains controversial. See for instance \cite{PhysRevLett.98.227202,Shao213,PhysRevB.80.180414,PhysRevLett.104.177201,PhysRevLett.107.110601,PhysRevLett.111.087203,PhysRevB.91.104411,PhysRevLett.115.267203,PhysRevLett.108.137201}. For the honeycomb lattice, which has only $\ZZZ_3$ site-rotation symmetry, there is no 't Hooft anomaly in the field theory and no LSM theorem for the lattice, so there is the possibility for an intermediate trivial gapped phase. Possible ground states for this phase were constructed in \cite{featurelesshoneycomb,jian-zalatel,Po_2017}, although it is not clear how to reach them from the $\mathbb{CP}^1$ model.

Applying the Thouless pump invariant to this study we learn some new conclusions, including for the honeycomb lattice. The $SO(3)$-vector scalar mass terms $m^2 n_i = N_i$ correspond to N\'eel polarizing fields, so we will refer to the gapped state obtained by giving these terms a large coefficient as the polarized state. As we have argued, in the 3-parameter phase diagram of the $N_i$, regardless of other local perturbations such as the precise value of $M^2$ or the strength of symmetry-allowed monopole operators, on a large $S^2$ there will be a Thouless pump for the unbroken monopole symmetry, in other words for the site-rotation symmetry $\ZZZ_p$, $p = 2, 3, 4,$ or 6.

The Thouless pump implies that the phase diagram must have some diabolical locus for small $\vec N$. For instance, if we begin in the N\'eel phase, where $SO(3)$ is spontaneously broken, then for any nonzero $\vec N$ the system is polarized, so the gapless N\'eel phase appears as just a single gapless point at $\vec N = 0$, described in the IR by an $S^2$ sigma model. For $M^2 \gg 0$ on the other hand, we will open up a VBS phase at $\vec N = 0$. There will be a phase transition at some nonzero $\vec N$ into the polarized state.


\begin{figure}
    \centering
    \includegraphics[width=0.3\textwidth]{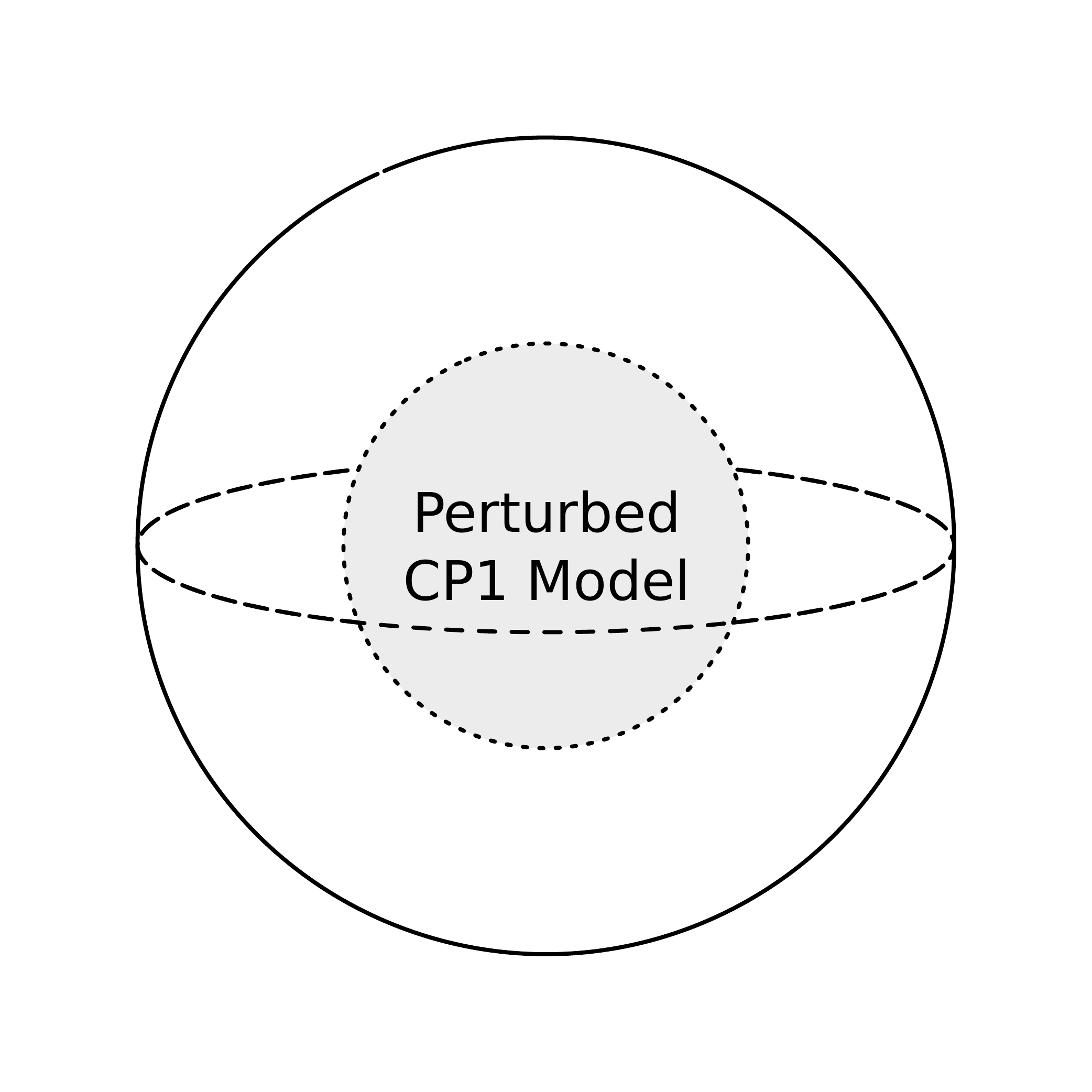}
    \caption{The three parameter phase diagram of the $\mathbb{CP}^1$ model with potential \eqref{eqn:scalarpotential}, at some value of $M^2$ and other perturbations not necessarily at the critical value, and where we have explicitly broken the $SO(3)$ symmetry by the three mass parameters $N_i = m^2 n_i$, which correspond to N\'eel polarizing fields in the N\'eel-VBS transition. The fate of the theory at the origin is unknown except in the large $M^2$ regions, but there is a Thouless pump for the $U(1)$ monopole charge on a sphere of large radius which protects a diabolical locus inside for all $M^2$. This is true even after breaking $U(1)$ to a cyclic subgroup of odd order, for which there is no 't Hooft anomaly or LSM constraint. Note that the phase diagram is spherically symmetric so long as the $N_i$ are the only $SO(3)$-breaking perturbations.}
    \label{figCP1}
\end{figure}

As we vary $M^2$ between these two regimes, there must always be a phase transition somewhere in the slice of the phase diagram corresponding to a fixed $M$. See Fig. \ref{figCP1}. This is true even on the honeycomb lattice, where one might go through an intermediate trivial phase for some $M$ and zero polarizing field. In that case, by the spherical symmetry of the phase diagram, this trivial phase will still be separated from the polarized phase at large $\vec N$ by a phase transition at some positive $|N|^2$. Near this phase transition, both nearby phases look trivial. This situation is fine-tuned in the sense that the $N_i$ are the only $SO(3)$-breaking perturbations. If we add other such perturbations, we can spoil the spherical symmetry of the phase diagram in Fig. \ref{figCP1}, resulting in a generic (although still nontrivial) diabolical locus.

If we include in the action a level $k$ Chern-Simons term for the dynamical $U(1)$ gauge field, then (\ref{eqn:uone2b}) implies that the Chern-Simons term produces a Hopf term with coefficient $k$ for the parameters $n_i$ \cite{Wilczek:1983cy}. As discussed in \cite{Wilczek:1983cy}, such term results in the skyrmion being a fermion or boson when $k$ is odd or even, in agreement with the spin of the monopole in the microscopic gauge theory. The computation can be generalized straightforwardly to $U(N)$ gauge theory coupled to two complex scalars.

Note that the 2+1d Thouless pump term coincides with that of two complex fermions in (\ref{eqn:3d2fcurrent}).
For $k=1$ the Hopf term coincides with the $\theta=\pi$ term for $n_i$ in the system of two complex fermion.
Thus the above computation provides a new consistency check for the duality between two complex fermions and $U(N)_1$ coupled to two complex Wison-Fisher scalars \cite{Aharony:2015mjs,Hsin:2016blu},\footnote{
The duality holds for $N\geq 2$ since otherwise the mass deformation in the scalar theory contains extra Goldstone mode \cite{Hsin:2016blu}.
}\footnote{We remark that the effective action discussed here cannot be removed by a well-defined local counterterm of the background field since it is not well-defined at $m^2=0$. Thus it must agree across the duality.
}
\begin{equation}
2\text{ Free Dirac fermions}\quad\longleftrightarrow\quad
U(N)_1+2\phi~.
\end{equation}
On the other hand, the theory with a more general Chern-Simons level $k$ may flow to interacting fixed points. In those models, there is a kind of Thouless pump, but the surrounding phase is a nontrivial TQFT, and the skyrmion binds a nontrivial anyon \cite{Freedhopf}.

It is also suspected that there is an emergent $SO(5)$ symmetry in this model, where the two monopole operators $M_1 \pm M_{-1}$ and the $n_i$ form an $SO(5)$ vector \cite{PhysRevLett.115.267203}.\footnote{
There are also results from conformal boostrap against a stable  critical point with $SO(5)$ symmetry in $(2+1)d$ \cite{Nakayama:2016jhq}.
} If we consider perturbations by these operators, the theory describes a diabolical locus in a 5-parameter phase diagram, protected by a WZW term of level 1 \cite{Tanaka_2005,Senthil_2006,Wang_2017}. The protection of this diabolical locus holds without assuming any symmetry at all! The WZW term can be derived using from the Thouless pump above, and vice versa, using equivariant cohomology techniques for computing anomalies of WZW theories \cite{Lapa_2017}.

The above computation can be generalized to other dimensions, with the 0-form symmetry for the Thouless pump invariant replaced by higher-form symmetry. We will give an example in section \ref{sec:uone2phi4d}.

\subsection{Gauge theories in 3+1d}

\subsubsection{$U(1)$ gauge theory with two scalars}\label{sec:uone2phi4d}

Consider a $U(1)$ gauge theory in 3+1d coupled to two complex scalars of charge one with an $SU(2)$-invariant potential (\ref{eqn:scalarpotential}).
The theory has a magnetic $U(1)$ 1-form symmetry with a current $j^{\mu\nu}=\frac{1}{2\pi}\epsilon^{\mu\nu\rho\sigma} \partial_\rho a_\sigma$ where $a$ is the gauge field \cite{Gaiotto:2014kfa}.
Then by repeating the computation in section \ref{sec:scalarThouless} we find that in the presence of a background 2-form gauge field $B^{(2)}$ for the $U(1)$ 1-form symmetry, the effective action for the family of theories away from the point $m^2=0$ has the following 3+1d Thouless pump invariant
\begin{equation}\label{eqn:Thoulesspump3+1dUone}
S_\text{top}=\frac{1}{8\pi}\int\epsilon^{\mu\nu\rho\sigma}\epsilon^{ijk} B^{(2)}_{\mu\nu} n_i\partial_\rho n_j\partial_\sigma n_k~ d^4x.
\end{equation}
The significance of this topological terms is as follows. For a fixed $m^2$, the parameter space of the family of theories we consider is $S^2$. Therefore it makes sense to consider skyrmionic strings: static topologically nontrivial configurations of the parameter fields $n^i$ which depend on two out of three coordinates on the space $\RRR^3$ and approach a constant at infinity. Such strings are classified by a skyrmion 1-form charge taking values in $\pi_2(S^2)=\ZZZ$. The term (\ref{eqn:Thoulesspump3+1dUone}) makes skyrmionic strings charged with respect to the 1-form $U(1)$ symmetry, so that their $U(1)$-charge is identified with the skyrmion charge. This protects the gapless locus at $m^2=0$ provided 1-form $U(1)$ symmetry remains unbroken.

The stablility of the $m^2=0$ point is also protected by an order 2 mixed 't Hooft anomaly between the magnetic 1-form symmetry and the $PSU(2)=SO(3)$ flavor symmetry.\footnote{
To see this, we can turn on background $SO(3)$ gauge field with nontrivial $w_2^{SO(3)}$ which is the obstruction to lifting the bundle to an $SU(2)$ bundle, then the Dirac quantization of the $U(1)$ gauge field $a$ is modified to be
\begin{equation}
\oint\frac{da}{2\pi}=\frac{1}{2}\oint w_2^{SO(3)}\;\text{mod }\mathbb{Z}~.
\end{equation}
Thus the coupling $\int B^{(2)}da/(2\pi)$ to background $B^{(2)}$ for the magnetic 1-form symmetry is no-longer 3+1d, and we find the theory lives on the boundary of the SPT phase
\begin{equation}
\pi\int_{5d} \frac{dB^{(2)}}{2\pi}\cup w_2^{SO(3)}~.
\end{equation}
}
However, the previous argument using the Thouless pump invariant does not rely on $SO(3)$ symmetry and thus is still valid for perturbations that break $SO(3)$. Moreover, such an anomaly becomes trivial when we consider two copies of the system.
The argument based on the effective action (\ref{eqn:Thoulesspump3+1dUone}) shows that neither the number of copies nor the $SO(3)$ flavor symmetry are essential for the stability of the phase transition at $m^2=0$.\footnote{
The same comment applies to the 2+1d theory discussed in Section \ref{sec:scalarThouless} whose discrete 't Hooft anomalies were  studied in \cite{Benini:2017dus}.}

\section{The Bulk-Boundary Correspondence for Spheres}\label{secbulkboundary}

We will show that given any boundary condition of a $d$-dimensional system with a higher Berry number on a parameter space with the topology of $S^{d+2}$, there is at least one parameter value where the boundary gap closes, indicating either gapless edge modes or an edge degeneracy. In the case that the parameter space is $\mathbb{R}^{d+2}\backslash \{0\}$, repeating the argument on successive spheres we find an arc connecting the origin to infinity, along which the boundary gap closes. Furthermore, if we consider varying parameters near the boundary, along a small $S^{d+1}$ which links the singularity, we find the boundary theory has a higher Berry number on the linking $S^{d+1}$. This by itself is not anomalous, but the fact that this $S^{d+1}$ lies in a larger parameter space where it is homotopically trivial requires the existence of the bulk. Compare Sections \ref{subsubsec1dthoulessinterfaces}.

\begin{figure}
    \centering
    \includegraphics[width=0.5\textwidth]{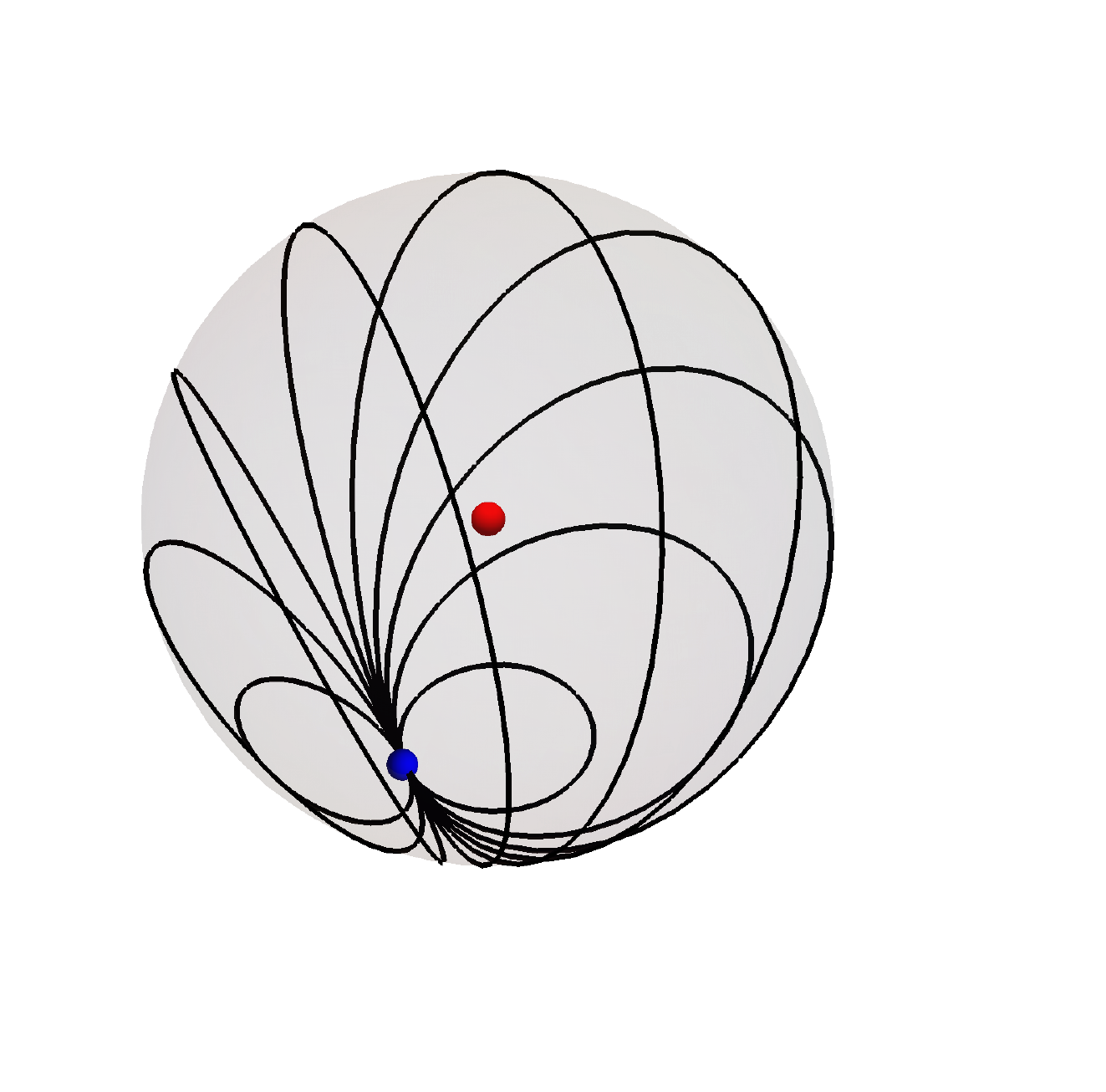}
    \caption{A 1-parameter family of backgrounds on spacetime $\RRR^{d+1}$ (or a ball $B^{d+1}$) useful for measuring the higher Berry number around a parameter space $S^{d+2}$ and for arguing the bulk-boundary correspondence. Here we have drawn the case $d = 0$, with each hoop depicting the image of the parameter map $\phi$. We fix the parameter values at spacetime infinity (or sufficiently near the boundary) at the blue point, where all hoops meet, and varying the parameters in a compact but large region so that over the 1-parameter family, the parameter values wind around the sphere once. Note that this family begins and ends at a background where the parameters are constant, at the blue point. Thus, the partition function in this process will return to itself. Assuming a uniform energy gap over the sphere implies that this partition function never vanishes, hence the winding number of the phase of the partition function over this process is a quantized topological invariant, which we identify with the higher Berry number by the effective action \eqref{eqn1dberry} and its higher-dimensional generalizations. The red point is a hypothetical diabolical point which sources the Berry curvature.}
    \label{fighoops}
\end{figure}

To see there must be a boundary singularity, we will show the partition function vanishes at a certain point by studying the winding number of its phase. Indeed, far from the boundary, but still over a wide-enough region so that we do not cause the bulk gap to close, we can cause the phase of the partition function to wind a number equal to the higher Berry number. We do this by choosing spatially-varying parameters which draw a ``hoop'' in parameter space, such that along a 1-parameter family of such hoops beginning and ending at a chosen configuration of constant parameter, we envelop the target space once. See Fig. \ref{fighoops}.

So long as the partition function does not vanish, this winding number is a quantized integer. The nonvanishing of the partition function is ensured by the uniform gap assumption. On the other hand, since we are on a spacetime with boundary, we can move the hoop region to the edge, absorbing it and unwinding the 1-parameter family so constructed. At the end of this, the winding number must be zero. Therefore, the partition function must vanish somewhere during this process, so the assumption of the uniform gap must be violated by the boundary.

\begin{figure}
    \centering
    \includegraphics[width=0.5\textwidth]{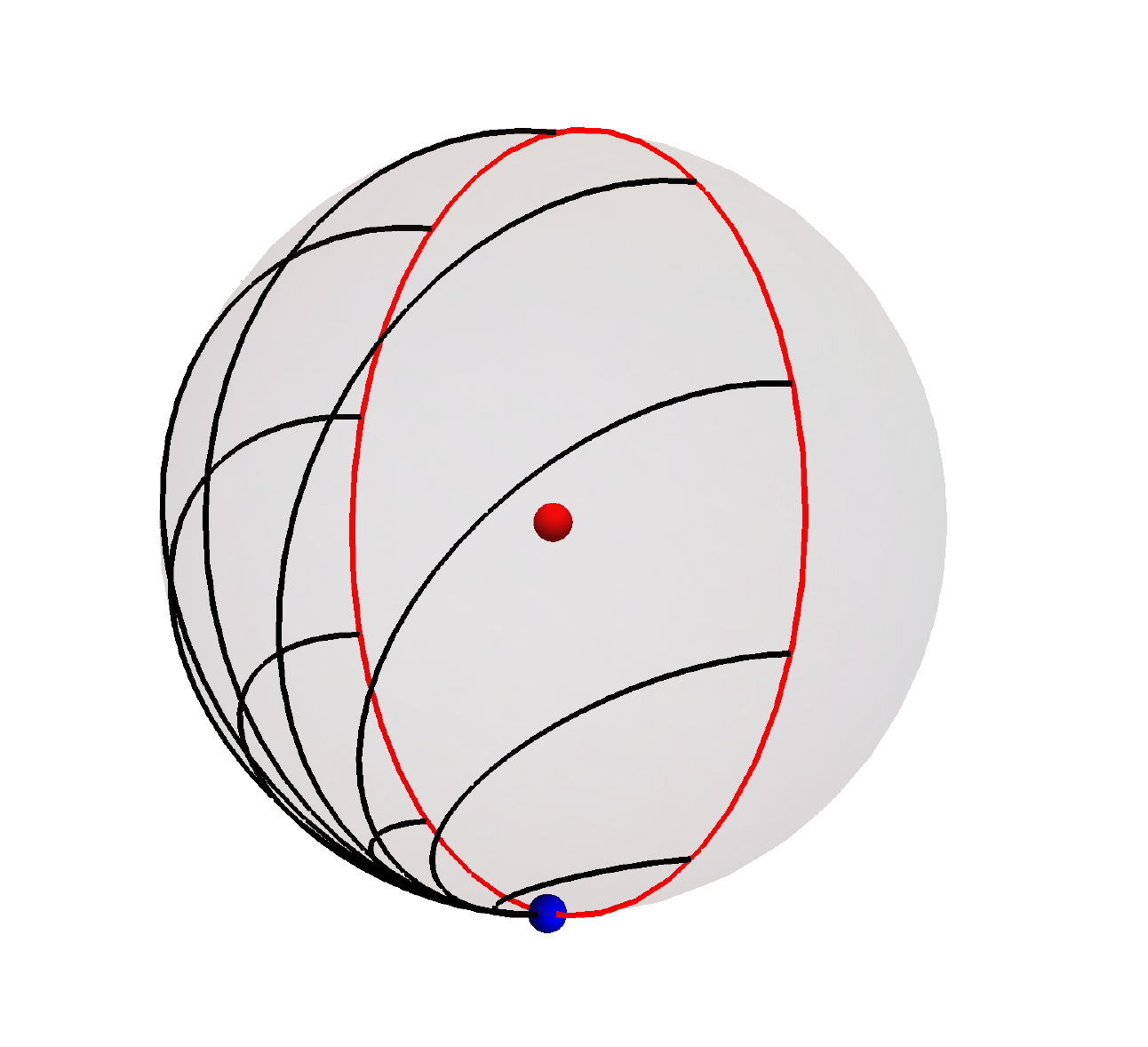}
    \caption{A 2-parameter family of backgrounds where the boundary is used to unwind the 1-parameter family in Fig. \ref{fighoops}, depicted at the point where the latter family is half-way ``absorbed" into the boundary. We have drawn the image of the parameter values at the boundary as a red circle. Because this family of families interpolates between a trivial family and one for which the phase of the partition function winds, the partition function must vanish for some value of the two parameters, which implies the gap must close at the boundary at that value. This occurs precisely where the red circle crosses a boundary diabolical point. We may define a boundary Berry number for such points. We find that the sum of these boundary Berry numbers equals the bulk Berry number.}
    \label{fighoopretract}
\end{figure}

We can perform this unwinding more systematically by using the boundary to retract each hoop. See Fig. \ref{fighoopretract}. This defines a 2-parameter family $\phi(u,v)$ of backgrounds. Let $u$ denote the parameter of the original family (which may be considered a circular parameter) and $v$ the parameter of the unwinding. We have argued there is some $u_*, v_*$ where the partition function vanishes, hence there is some parameter value $p_*$ in the image of $\phi(u_*,v_*)$ where the boundary is either degenerate or gapless. Typically there could be several such points.

In a small neighborhood of such a point, we may integrate out the bulk and consider $p_*$ as a diabolical point for the effective boundary theory. This implies that such points are isolated in the absence of extra symmetries. Furthermore, they are protected by a higher Berry number of the boundary, which we can compute as follows.

The special point $p_*$ is isolated so there is some $\epsilon > 0$ such that there are no other boundary diabolical points within the parameter values in the image of the strip $v_* - \epsilon < v < v_* + \epsilon$, $u \in S^1$. We form a 1-parameter family of backgrounds by composing the 1-parameter families $\phi(u,v_* + \epsilon/2)$ and $\phi(2\pi - u,v_*-\epsilon/2)$ (these are composable because all families begin and end at the blue point in Fig. \ref{fighoopretract}). This family is constructed so that the parameter values on the boundary wind once around the boundary diabolical point $p_*$. Further, because we are in the strip away from any other diabolical points, the partition function is nonvanishing in this family. Finally, the winding number jumps between the families $\phi(u,v_* + \epsilon/2)$ and $\phi(u,v_* - \epsilon/2)$, and we have composed them with different orientations in $u$, so the one so constructed has nonzero winding number. This winding number protects the boundary diabolical point by our usual arguments, and thus is very analogous to a bulk higher Berry number.

There is an important subtlety, however, which distinguishes this situation from cases where there is no bulk. Indeed, if we sum up the winding numbers defined above over all the boundary diabolical points, we get something nonzero---in fact one can readily see we get the bulk higher Berry number! This ``index theorem" is an aspect of anomaly in-flow for the higher Berry phase. Indeed, without the bulk, for a compact parameter space it is easy to see the total index of diabolical points (summing their higher Berry numbers) must be zero.

To put it another way, our 1-parameter family we defined above to encircle the boundary diabolical point $p_*$ is homotopic without changing the boundary values to one where the bulk parameter values lie only in the special strip. On the other hand we could've defined a 1-parameter family with the same boundary values but where the bulk parameter values lie \emph{outside} the strip, such that the two 1-parameter families are complementary, in that we can glue them together along the boundary to form something homotopic to our bulk 1-parameter family in Fig. \ref{fighoops}. Thus, while these 1-parameter families have precisely the same boundary parameters, their winding numbers differ by the bulk higher Berry number.

This confirms a point of view in \cite{Cordovaetal1}, which is that theories with anomalies in the space of coupling constants are typically presented with a noncompact parameter space, and while the ground state is trivially gapped at infinity, there is some winding that prevents one from compactifying the parameter space. Here we see that we expect such compactification is possible iff one introduces a bulk. We comment on the applications of these arguments to interfaces in Appendix \ref{appdimred}.


This discussion can be extended to systems with a global symmetry group $G$. We study families parameterized by $S^{d+1-k}$, $k \ge 0$, which are characterized by a generalized Thouless pump for which some $k$-spacetime-dimensional $G$-SPT $\Sigma_k$ is carried by the skyrmion. After compactifying along a $k$-dimensional test manifold with background gauge field for this SPT, we can reduce the detection of this SPT on the skyrmion by measuring the phase (or winding number thereof) of the sphere partition function. Introducing a boundary and using it to unwind the skyrmion as in Fig. \ref{fighoopretract}, we again find there is a boundary diabolical point where this partition function vanishes. If $d+1-k = 1$, at this point the boundary looks like a phase transition between $G$-SPTs. If $d + 1 - k > 1$, the boundary diabolical point is associated with a Thouless pump over an $S^{d-k}$. As before, the sum of the $G$-SPT classes associated with each of these diabolical points must be $\Sigma_k$.


\section{Classification of diabolical loci}\label{secgenremarks}

In this section we discuss a conjectural classification of higher Berry numbers and diabolical loci. Consider the space $\CM_d$ of infrared-trivial systems, that is, those with a unique, gapped ground state and no topological degrees of freedom. The connected components of this space (elements of $\pi_0 (\CM_d)$) are the short-range entangled (SRE) phases of matter. It is by now well accepted, although it remains unproven, that in the case of bosons (resp. fermions), $\pi_0 (\CM_d)$ can be expressed in terms of cobordism groups of oriented (resp. spin) closed manifolds \cite{kapustin2014symmetry,Kapustin_2015}. 

In fact, there is a much stronger conjecture, motivated by TQFT, which identifies the entire homotopy type of $\CM_d$ with that of a space in a certain cobordism spectrum \cite{freed2016reflection}. The fact that the homotopy type forms a loop-spectrum is very powerful, and allows us to immediately derive the classification of SRE phases with global symmetries ({\it i.e.} SPT phases) \cite{Kitaev_talk,Gaiotto_2016,Kapustin_2017,Xiong_2018,Gaiotto_2019,thorngren2018anomalies} as well as classify \emph{families} of SRE phases \cite{Thorngren_2018}, which is what we are interested in.

For instance, a basic consequence of the existence of a loop-spectrum for SRE phases is the isomorphism
\begin{equation}\pi_k (\CM_d) = \pi_0 (\CM_{d-k}),\end{equation}
which says that a family of infrared-trivial systems parametrized by $S^k$ can be characterized by a generalized Thouless pump which pumps a $d-k$-dimensional SRE phase to the boundary.

We see that if $d$ and $k$ are increased together, $d-k$ does not change, so for each $m = d-k$ there is a \emph{series} of families associated to an $m$-dimensional SRE phase. The Abanov-Wiegmann ``A" series in \eqref{eqnAWevenA}, \eqref{eqnAWoddA} for instance is associated to the 0+1d fermionic SRE ``phase'' characterized by the fact that the ground state has fermion number $1$ (or $1\, {\rm mod}\, 2$ if only the fermion parity is conserved.) This phase is the generator of
 $\pi_0 (\CM_1) = \Omega^1_{spin^c} = \ZZZ$ in the $U(1)$ conserving case or $\pi_0 (\CM_1) = \Omega^1_{spin} = \ZZZ_2$ in the case with just fermion parity. The theory studied in Section \ref{subsecaxions} with the axion coupling corresponds to $m = 2$, with $\pi_0(\CM_2) = \ZZZ$ generated by the $p+ip$ superconductor.
 
 The ``B" series in \eqref{eqnAWevenB}, \eqref{eqnAWoddB} corresponds $m=-1$.  Since $(-1)$-dimensional systems do not make sense, one cannot associate this series to an SRE phase. However, one can set $k=1$ and associate it to a non-contractible loop of SRE ``systems'' in $0+0$ dimensions. A ``system'' in zero dimensional space-time is characterized by its partition function, which is a complex number. For an SRE ``system'' this number is nonzero. The non-contractible family in question is the one where the phase of the partition function  winds once around the origin of the complex plane. Although this series has $m=-1$, geometrically it corresponds to the generator of the bordism group of spin 0-manifolds, which is $\ZZZ$. This degree shift is a consequence of Anderson duality applied to the bordism spectrum \cite{freed2016reflection}. With time reversal symmetry, the partition function is real, so there is no integer Berry number, but for $m=0$ there are two SRE ``phases'', the trivial one and the non-trivial one. The non-trivial one gives rise to a series of codimension $d+2$ diabolical loci of time-reversal-invariant systems in spatial dimension $d$. The one for $d=0$ was observed already by von Neumann and Wigner \cite{vN_Wigner} and corresponds to the holonomy of the Berry connection being $-1$. Geometrically this series arises from the generator of the bordism group of unoriented points, which is $\ZZZ_2$. It does not get shifted in degree by Anderson duality.

These considerations are the interacting analog of the work of Teo and Kane \cite{TeoKane} on topologically protected defects in gapped systems of free fermions in $d$ spatial dimensions. They argued that the classification of such defects depends only on $\delta=d-k$, where $k+1$ is the codimension of the defect. A defect of codimension $k+1$ in a system of free fermions can be created by making the system's parameters depend on the coordinates of $S^k$ which surrounds the defect. Therefore a topologically protected defect corresponds to a nontrivial element of $\pi_k(\CK_d)$, where $\CK_d$ is the space of gapped systems of free fermions in $d$ dimensions. Since the spaces $\CK_d$ form a loop spectrum too (the K-theory spectrum \cite{periodic_kitaev,periodic_fourauthors_short,periodic_fourauthors_long}), one finds again that $\pi_k(\CK_d)=\pi_0(\CK_{d-k}).$ Thus the classification of topologically protected defects depends only on $d-k$ and can be obtained from the periodic table of topological insulators and superconductors \cite{periodic_kitaev,periodic_fourauthors_short} by a shift $d\mapsto d-k$. In particular, Bott periodicity in K-theory implies that the classification of defects is periodic in $k$ with period $2$ or $8$ depending on the symmetry class.\footnote{Some of these K-theory invariants, such as the higher Chern numbers, have been interpreted in terms of curvatures of higher form connections on the Brillouin zone in \cite{Palumbo_2018,Palumbo_2019}.}

If one views $k+1$ as the codimension of a spatial defect, then $k$ cannot be larger than $d$. But if one views $k$ as the codimension of a diabolical point in the phase diagram, then $k$ can be arbitrarily large. Bott periodicity thus implies that in a $d$-dimensional system of free fermions the codimension of a diabolical point can be arbitrarily high. In contrast, in the interacting case there are no diabolical points of codimension greater than $d+3$. Interactions destabilize all diabolical points of higher codimension. An argument for this is sketched in Appendix \ref{appnofurther}.

For a general parameter space $P$, the classification implies that families of $d$-space-dimensional infrared-trivial systems parametrized by $P$ are classified by a certain cobordism group for closed $d+1$-manifolds equipped with a map to $P$, which we 
\begin{equation}\left\{ \substack{{\rm families\ of\ infrared-trivial\ systems} \\ {\rm of\ bosons\ or\ fermions\ parameterized\ by\ }$P$}\right\} \simeq \Omega^{d+1}_{\rm SO\ or\ Spin}(P).\end{equation}
There is a ``supercohomology" approximation to this classification which was described in \cite{Freed:2006mx}. In particular, if $P$ supports fermion parity Thouless pumps, it can affect the quantization of the WZW term to be fractional relative to the quantization on a sphere. If there is a global symmetry $G$, we study instead the cobordism group of $P \times BG$, with appropriate twists \cite{Kapustin_2015}. The group on the right hand side is interpreted as cobordism-invariant effective actions for a $d+1$-dimensional spacetime with a map to $P$, \textit{i.e.} a spatiotemporally-varying background.

The correspondence between families of theories and theories equipped with varying parameters, or parameters promoted to fields, is quite general, and has been used in many contexts in physics. There is mathematical evidence for this correspondence. For instance, it has been observed in \cite{Thorngren_2018} that the Baez-Dolan-Lurie cobordism hypothesis \cite{Baez_1995,Lurie_2008} implies an equivalence between families of TQFTs parametrized by a space $P$ and TQFTs for manifolds equipped with a map to $P$. In the general case, looking at the space of systems $\CM^{\mathcal{T}}_d$ whose IR limit is a $d$+1-dimensional TQFT $\mathcal{T}$, one expects for $0 < k < d+2$, $\pi_k (\CM_d^{\mathcal{T}})$ is the group of invertible $d-k$-dimensional topological defects in $\mathcal{T}$, $\pi_{d+2} (\CM_d^{\mathcal{T}}) = \ZZZ$, and higher homotopy groups vanish. This is relatively well-understood for $d \le 2$ \cite{Etingof_2010,Fuchs_2013}, and an understanding is slowly emerging in higher dimensions \cite{johnsonfreyd2020classification}.

Finally, there is the question of how these various spaces are embedded into the space ${\mathsf M}_d$ of all low-energy theories in $d$ space dimensions. This space may have several components, labelled by anomalies for instance. Our spaces $\CM_d$ are embedded in the component ${\mathsf M}^0_d$ of ``anomaly-free" low-energy theories. One might expect this space is homotopic to the space of local lattice Hamiltonian models, a convex space, hence that ${\mathsf M}^0_d$ is contractible. This would be satisfying, since it implies that every nontrivial homotopy group of $\CM_d$ (or any other stratum) corresponds to some diabolical point on a deeper stratum inside.

\emph{Note added:} at the same time as this paper, another paper on a closely related subject appeared on the arXiv \cite{sharon2020global}. That paper studied a phenomenon they dubbed ``vacuum crossing", wherein traversing a noncontractible loop in a region of parameter space with multiple vacua (e.g. in a spontaneous symmetry breaking phase), the system undergoes an adiabatic transformation from one ground state to another. If $M$ is a region of parameter space where the system has $n$ degenerate ground states, then this adiabatic transformation is encoded in a permutation representation $\pi_1(M) \to S_n$. If these $n$ degenerate ground states form a representation of some spontaneously broken symmetry $G$, then this permutation must be valued in the commutant of $G$. If they are associated with anyons, such as when considering ground states of a 2+1d TQFT on a torus, then this permutation representation must respect the braiding action.

One point of view which encompasses both vacuum crossing and the invariants discussed in this paper is to say that for any subset $M$ of parameters where the system flows to the same IR fixed point $\CT$, the \emph{homotopy type} of $M$ acts upon $\CT$ as a (higher) symmetry, so that for $0 < k < d+2$, any element of $\pi_k(M)$ is associated with a symmetry defect of $\CT$ (which may be spontaneously broken), while $\pi_{d+2}(M)$ is associated with higher Berry phases. If $M$ is homotopy equivalent to $S^1$, this is the permutation representation mentioned above. If $M$ is homotopy equivalent to $S^{d+2}$, where $d$ is the space dimension, then this is the higher Berry phase. Taking $M = \CM^T$ the space of all theories which flow to $\CT$, there is a universal such action. It would be very interesting to try to understand this in detail. We thank Adar Sharon for a discussion about this and his paper.

\section*{ACKNOWLEDGEMENT}

We thank Dominic Else, Tobias Holder, Raquel Queiroz, Nathan Seiberg, and Ruben Verresen for discussions.  A. K. is grateful to Lev Spodyneiko for a collaboration on a closely related project. The work is supported by the U.S. Department of Energy, Office of Science, Office of High Energy Physics, under Award Number DE-SC0011632, and by the Simons Foundation through the Simons Investigator Award.

\newpage
\bibliographystyle{apsrev4-1}
\bibliography{bib.bib}

\onecolumngrid
\appendix

\section{Topological actions for free fermions in $d$ dimensions}\label{appfreeferm}

\subsection{ Wess-Zumino-Witten terms}

Consider $N$ free massive fermions in $d+1$ spacetime dimension. The mass term in the Lagrangian has the form $m\bar\psi^aP_{ab}\psi^b~ $, where we parameterized the mass matrix $\mathbf{m}=m\mathbf{P}$ by an overall scale $m>0$ and a dimensionless matrix $\mathbf{P}$ with entries $P_{ab}$. 

In odd spacetime dimensions  Lorenz-invariance forces the mass matrix to act trivially on the spinor indices. Unitarity then requires $\mathbf P$ to be hermitian. The physical masses of fermions are $m$ times the absolute values of the eigenvalues of this matrix. Thus the model describes massive fermions if and only if ${\mathbf P}$ non-degenerate. If we have $N$ Dirac fermions, we can diagonalize ${\mathbf P}$ with a $U(N)$ transformation, and all eigenvalues are non-zero. We can deform them all to $\pm 1$ while keeping them non-zero. Hence we can assume that ${\mathbf P}^2=1$. 

In even spacetime dimensions we can write $\mathbf{P}=\mathbf{P}_1+i\gamma^{0 \cdots d}\mathbf{P}_2$ with hermitian $\mathbf{P}_1,\mathbf{P}_2$ which act trivially on the spinor indices, and with $\gamma^{0\cdots d}$ the chirality operator. In the cases of interest to us, these two matrices commute (in fact, $\mathbf P_1$ is always a scalar matrix). For $N$ flavors of fermions a  $U(N)$ rotation $\psi\mapsto U\psi$ can diagonalize both $\mathbf{P}_1$ and ${\mathbf P}_2$. The physical masses of fermions squared are $m^2$ times the eigenvalues of ${\mathbf P}_1^2+{\mathbf P}_2^2$. If we deform all masses to be $m$, the matrix ${\mathbf P}$ satisfies  $\mathbf{P}^\dagger {\mathbf P}=1$. The conclusion is that for all $d$ we can assume $\mathbf{P}^\dagger {\mathbf P}=1$.

The variation of the effective action with respect to the parameter $\mathbf{m}$ is
\begin{equation}
  \delta S_\text{eff}=\text{Tr }\delta\mathbf{m} (D^\dag D)^{-1}D^\dag~,
\end{equation}
where $D=i\slashed{\partial}+i\mathbf{m}$, and $D^\dag D=-\partial^2+\mathbf{m}^\dag \mathbf{m}-\slashed{\partial}\mathbf{m}$.
The topological term in the effective action is
\begin{align}\label{eqn:WZ00}
  \delta S_\text{eff}
   & =-i\text{Tr }m^{d+3}\delta\mathbf{P}(-\partial^2+m^2)^{-1} \cdot \left((-\partial^2+m^2)^{-1}  \slashed{\partial}\mathbf{P}\right)^{d+1}
  \mathbf{P}^\dag
  \cr
   & =-i\theta_0\int \text{Tr }\left(\delta\mathbf{P}(\slashed{\partial}\mathbf{P})^{d+1} \mathbf{P}^\dag\right)~,
\end{align}
where in the last line we used $\mathbf{P}^\dag\mathbf{P}=1$, and
\begin{equation}
\theta_0=\int\frac{d^{d+1}p}{(2\pi)^{d+1}}\frac{m^{d+3}}{(p^2+m^2)^{d+2}}=i\frac{\pi^{(d+1)/2}}{(2\pi)^{d+1}}\frac{\Gamma(3/2+d/2)}{\Gamma(d+2)}
=\frac{i}{4^{d+1}\pi^{d/2}\Gamma\left(\frac{d+2}{2}\right)}~.
\end{equation}
We can perform the trace over spinor indices using the identities
\begin{align}\label{eqn:traceidn}
  \text{odd }d:\quad &
  \text{Tr }\left(\gamma^{i_1}\gamma^{i_2}\cdots\gamma^{i_{d+1}}\gamma^{0\cdots d}\right)=i(-2i)^{[(d+1)/2]} \epsilon_{i_1,i_2,\cdots,i_{d+1}}\cr
  \text{even }d:\quad  &
  \text{Tr }\left( \gamma^{i_1}\gamma^{i_2}\cdots\gamma^{i_{d+1}}\right)=i(-2i)^{[(d+1)/2]} \epsilon_{i_1,i_2,\cdots,i_{d+1}}~,
\end{align}
which can be derived from cyclic property of trace and anti-commutation relation of the gamma matrices.
In odd spacetime dimension (even $d$) the result is the following Wess-Zumino-Witten  term:
\begin{equation}
  \theta_0{(-2i)^{d/2}\over d+2}
  \int \text{Tr }\left(\mathbf{P}(d\mathbf{P})^{d+2}\right)~.
\end{equation}
In even dimensions one can perform a similar computation with $\mathbf{P}=\mathbf{P}_1+i\gamma^{0 \cdots d}\mathbf{P}_2$. We will not give a closed form expression here.

\subsection{Topological terms with a background gauge field}

One can perform a similar computation with an additional background gauge field $A$ for an  internal Lie group symmetry $G$. We will assume the mass matrix is invariant under $G$ symmetry.
The expansion of $(D^\dag D)^{-1}$ is modified to be
\begin{equation}
   (D^\dag D)^{-1}
    =
  \sum_{s} 
  (-\partial^2+m^2)^{-s-1}\sum_r\left(-i\slashed{\partial}\slashed{A}\right)^r (m\slashed{\partial}\mathbf{P})^{s-r}+\cdots~,
\end{equation}
where the last term should be understood as a sum over different orderings of $\slashed{\partial}\slashed{A}$ and $\slashed{\partial}\mathbf{P}$.
The terms contributing to a $(d+1)$-form in the effective action correspond to $2r+s-r=d+1$, namely $s=d+1-r$.

The variation of the effective action thus contains
\begin{equation}\label{eqn:WZA00}
\sum_r(-i)^{r+1}\theta_{r,d}  \text{Tr }\delta\mathbf{P}\left( (\slashed{\partial}\slashed{A})^r (\slashed{\partial}\mathbf{P})^{d+1-2r}\right)\mathbf{P}^\dag~,
\end{equation}
where
\begin{equation}
 \theta_{r,d}
=\int\frac{d^{d+1}k}{(2\pi)^{d+1}}\frac{m^{d-2r+3}}{(k^2+m^2)^{d-r+2}}
=i\frac{\pi^{(d+1)/2}}{(2\pi)^{d+1}}\frac{\Gamma(3/2+d/2-r)}{\Gamma(d+2-r)}
  ~.
\end{equation}

\subsubsection{Odd dimension}

In odd spacetime dimensions (even $d$) we can use the trace identity (\ref{eqn:traceidn}) to obtain the following topological  terms in the partition function:
\begin{equation}
  \exp\left(i\int_{M_{d+2}} \text{Tr }e^{F/(2\pi)}\chi(\mathbf{P})\right)~,
\end{equation}
where $\partial M_{d+2}=M_{d+1}$, and
\begin{equation}
  \chi(\mathbf{P})\equiv\sum_s
 \frac{\sqrt{\pi}i^{s+d/2}}{(2\pi)^{s-d/2}}\frac{\Gamma(s-d/2+1/2)}{(2s-d)!}
\text{Tr}\left((d\mathbf{P})^{2s-d}\mathbf{P}\right)~.
\end{equation}

\subsubsection{Even dimension}

In even spacetime dimensions (odd $d$), we express the mass matrix as $\mathbf{P}=\mathbf{P}_1+i\gamma^{0\cdots d}\mathbf{P}_2$, where $\mathbf{P}_1,\mathbf{P}_2$ act only on the flavor indices. 
We can evaluate the trace in the variation of the effective action by applying the identity (\ref{eqn:traceidn}) that requires a single overall $\gamma^{0\cdots  d}$.
Consider the rightmost $\slashed{\partial}\mathbf{P}$ in (\ref{eqn:WZA00}) that contributes $\gamma^{0\cdots d}$
\begin{equation}
  \text{Tr }\left(\mathbf{P}^\dag\delta\mathbf{P}\cdots (i\slashed{\partial}\gamma^{d+2}\mathbf{P}_2)f_k\right)
\end{equation}
where $f_k$ is a product of powers of $(\slashed{\partial}\slashed{A})$ and $(\slashed{\partial}\mathbf{P}_1)$ that has a total of $k$ gamma matrices. Moving $\gamma^{0\cdots d}$ to the right end introduces $(-1)^k$. 
We repeat this procedure for the next one to the left contributing $\gamma^{0\cdots d}$ to the effective action until all $\gamma^{0\cdots d}$ are moved to the right. If the number of $\gamma^{0\cdots d}$ is $(2j+1)$, there is additional factor $(-1)^j$ from bringing $\gamma^{0\cdots d}$ through $\slashed{\partial}\mathbf{P}_2$.
Applying the trace identity (\ref{eqn:traceidn}) erases the gamma matrices and $f_k$ becomes a $k$-form. Then we move $d\mathbf{P}_2$ across the $k$-form to the right end, which introduces another sign that compensates $(-1)^k$. This gives
\begin{equation}
  \exp\left( i\int_{M_{d+2}} \text{Tr }e^{F/(2\pi)}\chi(\mathbf{P})\right)~,
\end{equation}
where the variation of $\int\chi(\mathbf{P})$ gives the boundary terms
\begin{align}
    \delta\chi(\mathbf{P})\equiv \sum_s&
  \frac{(s-(d+1)/2)!}{(2\pi)^{s-(d+1)/2}}i^{s-{d+1\over 2}}
    \left\{
  \sum_{j=0}^{[s-(d+1)/2]}\frac{1}{(2j)!(2s-d-2j-1)!}
 \text{Tr }\left(\mathbf{P}_1^\dag\delta \mathbf{P}_2-\mathbf{P}_2^\dag\delta\mathbf{P}_1\right)
  (d\mathbf{P}_1)^{2s-d-1-2j}(d\mathbf{P}_2)^{2j}
  \right.
  \cr
  & \left.
  +\sum_{j=0}^{[s-d/2-1]}
  \frac{1}{(2j+1)!(2s-d-2j-2)!}
  \text{Tr }\left(\mathbf{P}_2^\dag\delta\mathbf{P}_2+\mathbf{P}_1^\dag\delta\mathbf{P}_1\right)
  (d\mathbf{P}_1)^{2s-d-2j-2}(d\mathbf{P}_2)^{2j+1}
  \right\}
  ~.
\end{align}

For instance, in even spacetime dimensions $(d+1)$ the effective action of a single  massive fermion with $P_1=\cos \alpha,P_2=\sin\alpha$ contains the following Wess-Zumino term:
\begin{equation}
  i\frac{1}{(2\pi)^{(d+1)/2}((d+1)/2)!}\int_{M_{d+2}} d \alpha\;\text{Tr } F^{(d+1)/2}~.
\end{equation}
This is exactly as expected from the Atiyah-Patodi-Singer index theorem \cite{atiyah_patodi_singer_1975} that relates the $\eta$-invariant of the Dirac operator with $\text{Tr }e^{F/2\pi}$.

\section{Luttinger Liquid Deformations of the 1d Thouless Pump}\label{appluttingerthouless}

Both the Thouless pump and the diabolical point considered in Section \ref{subseccomplexferms1d} admit a simple description in terms of a compact boson. Let $\theta$, $\phi$ be $2\pi$-periodic dual fields describing a 1+1d compact boson. In our convention at radius $R$, the vertex operator $\exp (in\theta + iw\phi)$ has dimensions
\[(h,\bar h) = \left(\frac{1}{2}(n/R + Rw/2)^2, \frac{1}{2} (n/R - Rw/2)^2\right).\]
The free fermion theory is equivalent to $R = 2$, with charge conservation corresponding to the $U(1)$ shift symmetry of $\theta$.

There are two $U(1)$-invariant and Lorenz-invariant relevant operators, namely $\cos \phi$ and $\sin  \phi$. These correspond to the two mass parameters of the Dirac fermion. The axial $U(1)$ symmetry acts as the shift symmetry of $\phi$. If we turn on some linear combination of these mass terms, $\phi$ becomes fixed in the ground state due to a periodic potential. While all these ground states look trivial on their own, there is nevertheless a winding as we vary the argument of the mass around the origin, witnessed by the fixed value of $\phi$ winding around. Because $\partial_x \phi$ is the current for the $\theta$-shift symmetry, our $U(1)$, we immediately derive the topological current of Section \ref{subseccomplexferms1d}.

Suppose we deform the family of theories by keeping $R=2$ for sufficiently large $|M|$, but making it larger near $M=0$. If at some $|M|$ the radius $R$ exceeds $2\sqrt 2$, both $\cos\phi$ and $\sin\phi$ become irrelevant and the model flows to a Berezisnky-Kosterlitz-Thouless (BKT) phase with an algebraic decay of correlators. Thus the diabolical point gets resolved into an island of the BKT phase. On the boundary of this island the model undergoes a phase transition from the gapless BKT phase to the trivial gapped phase. 

\begin{figure}
    \centering
    \includegraphics[width=0.3\textwidth]{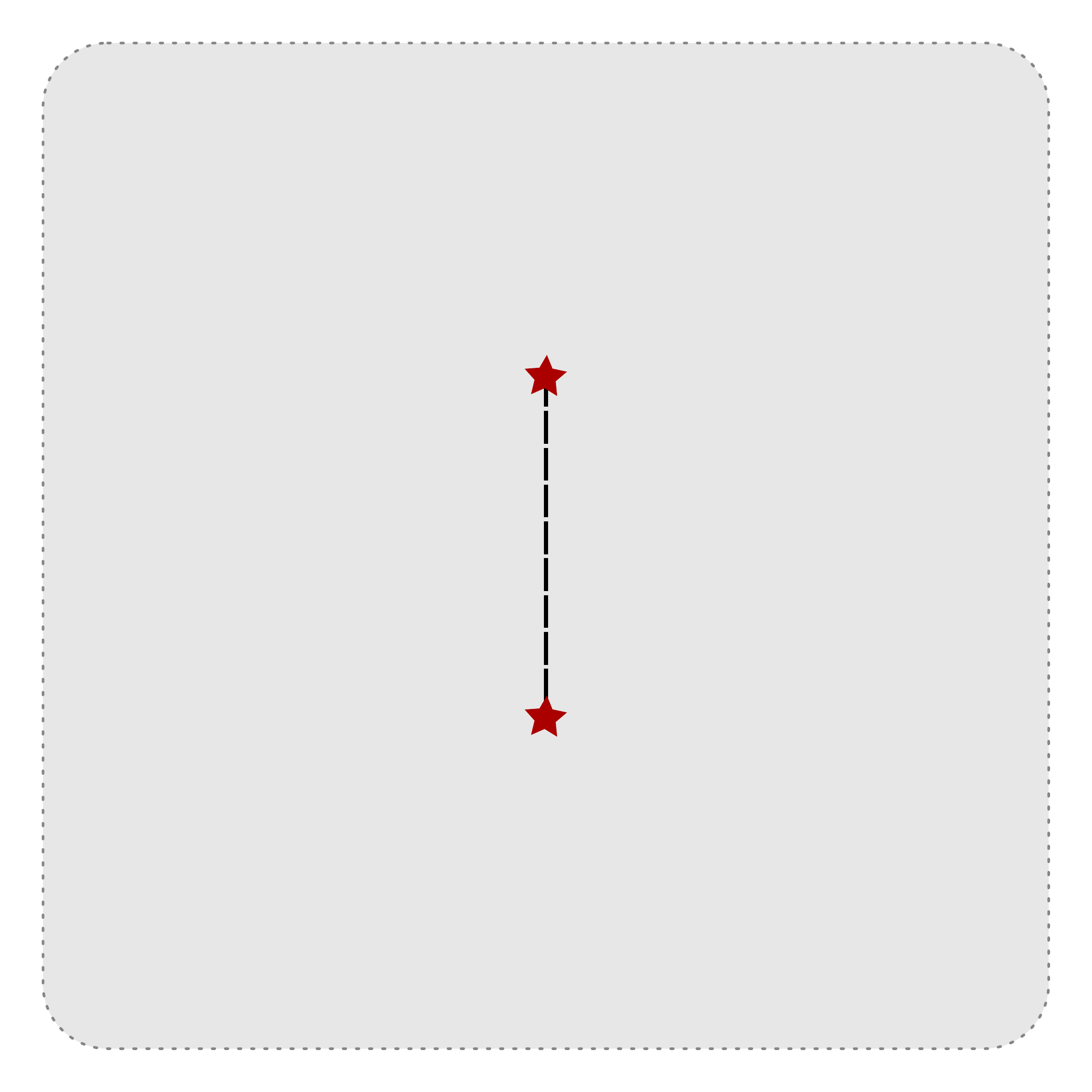}
    \caption{A first order line ending at two Ising critical points, protected by a $U(1)$ symmetry with a Thouless pump invariant encircling the diabolical locus (the interval and its endpoints). Without the $U(1)$ symmetry, the two points can come together and annihilate, leaving a phase diagram which is trivially gapped everywhere. With $U(1)$, however, they can only merge into a gapless Luttinger liquid point. There appears to be no way to rid the phase diagram of gapless points by $U(1)$-symmetric perturbations. The same conclusions hold even after breaking $U(1)$ down to just the fermion parity $\ZZZ_2$ subgroup.}
    \label{figfirstorderline}
\end{figure}

Alternatively, one can make $R$ smaller than $2$.
As we tune $R$ past the self-dual radius $R = \sqrt{2}$, two new operators become relevant: $\cos 2\phi$ and $\sin 2\phi$. In a generic 2-parameter phase diagram, this will cause the gapless Luttinger liquid point at zero mass to become destabilized. To see what happens, without loss of generality we may add a uniform perturbation $\cos 2\phi$ to the theory at each point in the 2-parameter phase diagram. Along the $\cos \phi$ axis of the phase diagram, there is an accidental charge conjugation symmetry
\[C:\begin{cases}\phi \mapsto -\phi \\ \theta \mapsto -\theta\end{cases}.\]
Near the origin, where $\cos 2\phi$ dominates $\pm\cos \phi$, $C$ is spontaneously broken, leading to a first order line. By studying the competition between these two potentials, we find this first order line ends at two Ising critical points at $\cos 2\phi \pm 4 \cos \phi$. This is depicted in Fig. \ref{figfirstorderline}.

If we are willing to break the $U(1)$ symmetry, then we can add $\cos 2\theta$ ($\cos \theta$ violates fermion parity and is not allowed). This creates a similar phase diagram as in the figure, with a first order line along the $C$-invariant axis ending in two Ising critical points.

If one studies this theory as describing a system of bosons, meaning we can also break the $\ZZZ_2$ shift symmetry of $\phi$, then we can consider the 4-parameter phase diagram of $\cos \phi$, $\sin \phi$, $\cos \theta$, $\sin \theta$. These operators form an $SO(4)$ vector at the $SU(2)_1$ radius $R = \sqrt{2}$. This implies that in that phase diagram the gapless point is isolated at the origin. The $SU(2)_1$ WZW term means that the higher Berry number is 1 for the 3-sphere which links the origin. See also \cite{haldaneNLSM}.

\section{Dimensional Reduction for Interfaces}\label{appdimred}

\subsection{Interfaces for the Abanov-Wiegmann Series}\label{appdimredAW}

The Abanov-Wiegmann ``B" series model in an odd space dimension $d$ is defined  as a free theory of  $2^{(d+1)/2}$ flavors of $2^{(d+1)/2}$-component complex fermions. The action has a $Spin(d+2)$ flavor symmetry, and the fermions transform in its spinor representation. If we denote the generators of the $(d+2)$-dimensional Clifford algebra  $\Gamma^a$, where $a = 1,\ldots,d+2$, the Lagrangian has the form
\begin{equation}\label{eqnAWevenB}
\CL_d^{\rm odd} = -i \bar \psi_j \gamma^\mu\partial_\mu \psi_j - i M^0 \bar \psi_j \psi_j - M^a \bar \psi_j \gamma^{0\cdots d} \Gamma^a_{jk} \psi_k,
\end{equation}
where $\gamma^{0\cdots d}$ is the chirality operator. Similarly, in an even space dimension $d$ we use $2^{d/2+1}$ flavors of $2^{d/2}$-component complex fermions transforming in the spinor representation of $Spin(d+3)$, with Clifford algebra generators denoted $\Gamma^a$, $a = 1, \ldots, d+3$ and the Lagrangian 
\begin{equation}\label{eqnAWoddB}
\CL_d^{\rm even} = -i \bar \psi_j \gamma^\mu \partial_\mu \psi_j - i M^a \bar \psi_j \Gamma^a_{jk} \psi_k.
\end{equation}
This parameterization is related to the one in \eqref{AWmodelB2d} by $n_i \sin \nu = M^i$, $i = 1,\ldots,d+2$, $m \cos \nu = M^{d+3}$.

Each of these models has a parameter space $\RRR^{d+3}$ spanned by the $M$'s, with a diabolical point at the origin and massive elsewhere, giving rise to a WZW term of level 1 for the unit sphere $S^{d+2}\subset \RRR^{d+3}$. 

We will show this by studying interfaces in the mass parameter space, which we realize as paths.
First we consider the even dimensional case $d = 2n$, $n > 0$, with $2^{n+1}$ complex fermions transforming under the flavor group $Spin(d+3)$ with the action \eqref{eqnAWevenB}. We can choose a basis where one of the $Spin(d+3)$ generators, which we take to be $\Gamma^{d+3}$, is of the diagonal form
\begin{equation}\label{eqndiagonalform}
    \begin{bmatrix} \mathbbm{1}_{2^n \times 2^n} && 0 \\ 0 && -\mathbbm{1}_{2^n \times 2^n} \end{bmatrix}  = \sigma^z \otimes \mathbbm{1}_{2^n \times 2^n},
\end{equation}
while $\Gamma^{0\cdots d}$ is of the form
\begin{equation}\label{eqnoffdiagonalform}
    \begin{bmatrix} 0 && \mathbbm{1}_{2^n \times 2^n} \\ \mathbbm{1}_{2^n \times 2^n} && 0 \end{bmatrix}  = \sigma^x \otimes \mathbbm{1}_{2^n \times 2^n},
\end{equation}
where $\sigma^x, \sigma^y, \sigma^z$ are the usual Pauli spin matrices. In this basis, $M^{d+3} \bar \psi \Gamma^{d+3} \psi$ looks like a diagonal mass term. We consider a Jackiw-Rebbi-like domain wall setup where $M^{d+3}$ varies along the coordinate $x_{d}$ from $-1$ for sufficiently negative $x_{d}$ to 1 for sufficiently positive $x_{d}$. As is well-known, each of the $2^{n+1}$ fermions will contribute a massless chiral mode localized on the wall. With $\Gamma^{d+3}$ as above, we find normalizability constrains the zero modes to satisfy
\begin{equation}\gamma^d \psi_j = \psi_j \qquad 1 \le j \le 2^n\end{equation}
\begin{equation}\gamma^d \psi_j = -\psi_j \qquad 2^n+1 \le j \le 2^{n+1}.\end{equation}

The resulting low-energy theory thus consists of $2^n$ complex \emph{nonchiral} fermions localized on the wall we describe as follows. If we choose a basis for the spinor labels where $\gamma^d$ has the diagonal form \eqref{eqndiagonalform}, we have
\begin{equation}\psi_j = \begin{bmatrix} \psi_j^+ \\ \psi_j^-\end{bmatrix},\end{equation}
where $\psi_j^\pm$ is a $2^n$-component spinor. We define the $2^n$ spinors on the wall (which have the same number of components as the $\psi$ but there are half as many of them) by
\begin{equation}\chi_j = \begin{bmatrix} \psi_j^+ \\ \psi_{2^n+j}^-\end{bmatrix}, \qquad \chi_j^+ = \psi_j^+ \qquad \chi_j^- = \psi_{2^n+j}^- \qquad 1 \le j \le 2^n.\end{equation}
We observe that $\Gamma^{d+3}$ acts the same as $\gamma^{d}$ on these spinors. Further, the interaction in \eqref{eqnAWevenB}
\begin{equation}M^a \bar \psi_j \Gamma^a_{jk} \psi_k\end{equation}
couples $\psi_j^\pm$'s of both opposite $\gamma^{d}$ and $\Gamma^{d+3}$. It restricts to an interaction among the $\chi_j$. To express this interaction, we first observe that in our chosen basis, with $\Gamma^{d+2}$ of the form \eqref{eqnoffdiagonalform},
\begin{equation}M^{d+2} \bar \psi_j \Gamma^{0\cdots d}_{jk} \psi_k = M^{d+2} \left( (\chi_j^+)^\dagger \chi_j^- + (\chi_j^-)^\dagger \chi_j^+\right)\end{equation}
is the diagonal mass term. The rest of the generators of $(d+3)$-dimensional Clifford algebra can be written in a basis compatible with \eqref{eqndiagonalform}, \eqref{eqnoffdiagonalform} by letting 
\begin{equation}\Gamma^a = \sigma^y \otimes \tilde \Gamma^a,\end{equation}
where $\tilde \Gamma^a$, $a = 1,\ldots,d+1$ generate a $(d+1)$-dimensional Clifford algebra. We find that the terms associated to $M^a$ over this range of $a$ thus take the form
\begin{equation}in^a (\chi_j^+)^\dagger \tilde \Gamma_{jk}^a \chi_k^- - i M^a (\chi_j^-)^\dagger \tilde \Gamma_{jk}^a \chi_k^+ = i M^a \bar \chi \gamma^{d} \tilde \Gamma^a \chi.\end{equation}
Summarizing, we have found the effective Lagrangian 
\begin{equation}
-i \bar \chi_j \gamma^\mu \partial_\mu \chi_j - M^{d+2} \bar \chi_j \chi_j - i M^a \bar \chi_j \gamma^{d} \tilde \Gamma^a_{jk} \chi_k, \qquad a = 1, \ldots, d+1\end{equation}
among $2^n$ complex fermions along the wall. We recognize this is the same as $\CL_{d-1}^{\rm odd}$ in \eqref{eqnAWevenB} after identifying $\gamma^{0\cdots (d-1)} =\gamma^d$.

\begin{figure}
    \centering
    \includegraphics[width=8cm]{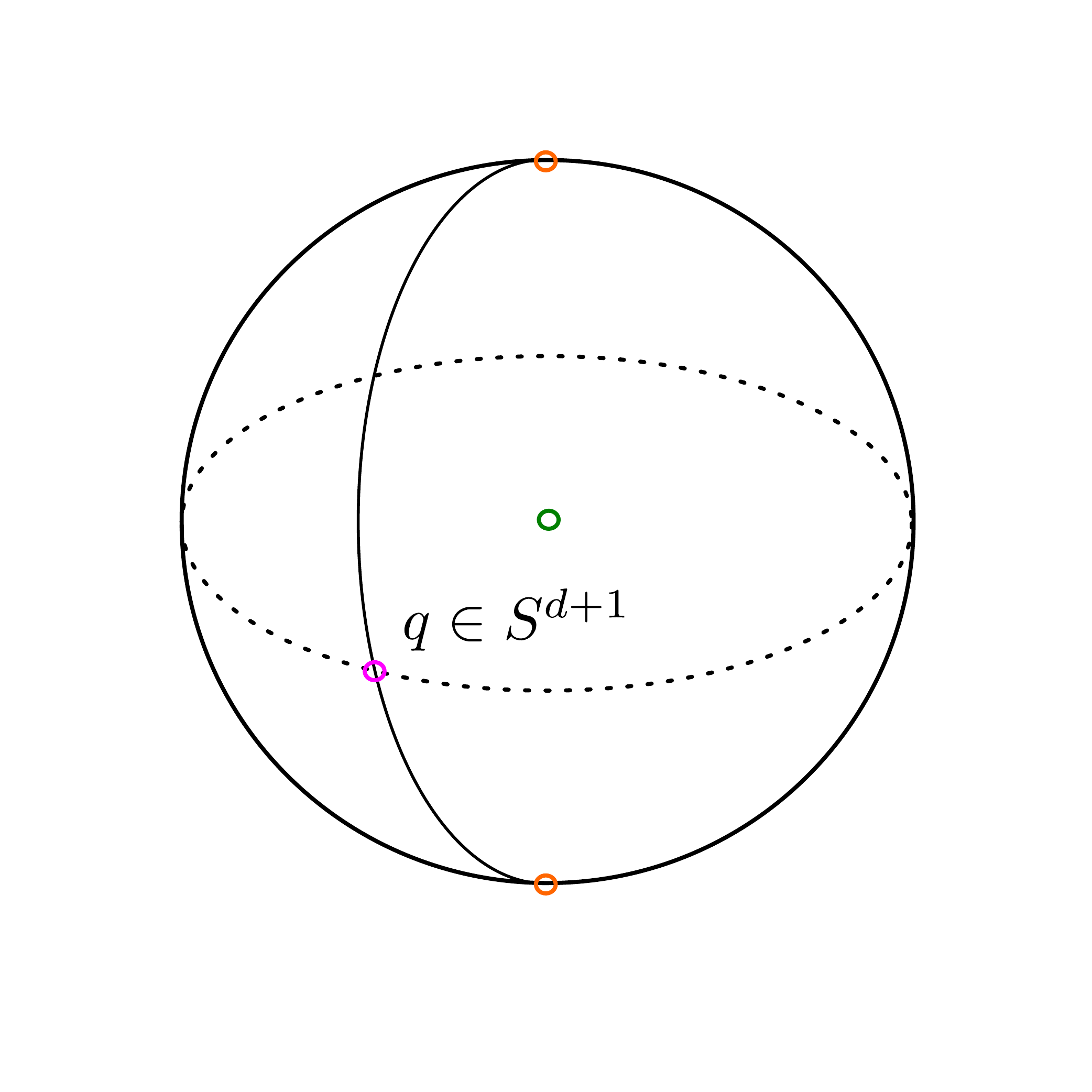}
    \caption{The spatially-varying parameters in $S^{d+2}$ for the domain wall configurations trace out a great semicircle (black) connecting the north and south poles (orange) and are parametrized by a point $q \in S^{d+1}$ (pink), where the semicircle intersects the equatorial $(d+1)$-sphere (dotted). The winding number of a family of such profiles over $S^{d+2}$ equals the winding number of the intersection points $q$ over $S^{d+1}$, giving an equality between the Berry number of the $d$-dimensional theory and its domain wall.}
    \label{figdimred}
\end{figure}

In this case there is a simple relationship between the Berry number of the $d$-dimensional theory parametrized by $S^{d+2}$ and the $d-1$-dimensional theory on the domain wall, parametrized by $S^{d+1}$. Indeed, the theory on the wall with given parameter values $M^a$, $a = 1,\ldots, d+2$ is defined from the $d$-dimensional theory with a spatially-varying parameters $M^a$ such that along the $x_{d}$ coordinate, the system parameters draw a great semicircle from the south pole $M^{d+3} = -1$ to the north pole $M^{d+3} = 1$. If the system parameters of $d-1$-dimensional theory wind once over the equatorial $S^{d+1}$, we see that the semicircular arcs also wind once over the whole $S^{d+2}$. Thus, the Berry numbers of the two theories are equal. See Fig. \ref{figdimred}.

Now we consider the odd dimensional case $d = 2n-1$, with $2^n$-many $2^n$-component complex fermions transforming in the flavor group $Spin(d+2)$ with generators $\Gamma^a$, $a = 1,\ldots, d+2$. We again consider the Jackiw-Rebbi problem with a spatially-varying mass $m(x_{d})$ which goes from $-1$ to $1$ over a region near $x_{d} = 0$. We find $2^n$ zero modes, one for each fermion, satisfying
\begin{equation}\gamma^{d} \psi_j = \psi_j.\end{equation}
We choose a basis for the spinor space where $\gamma^{d}$ takes the form $\sigma^z \otimes \mathbbm{1}$ of \eqref{eqndiagonalform}, so that
\begin{equation}\psi_j = \begin{bmatrix} \psi_j^+ \\ \psi_j^-\end{bmatrix}\end{equation}
and the constraint $\gamma^d \psi_j = \psi_j$ sets $\psi_j^-$ to zero. We define the $2^n$ domain wall fermions by
\begin{equation}\chi_j = \psi_j^+.\end{equation}
Note these have half as many components as the $\psi_j$'s.

Recall that the chirality operator is
\begin{equation}\gamma^{0\cdots d} = i^{(d-1)/2}\gamma^0 \cdots \gamma^{d}.\end{equation}
When this acts on $\chi_j$ it acts as
\begin{equation}i^{(d-1)/2} \gamma^0 \cdots \gamma^{d-1}\end{equation}
which is identically 1 on this subspace. Thus, our interaction simplifies to
\begin{equation}M^a \bar\chi_j \Gamma^a_{jk} \chi_k.\end{equation}
We thus obtain the theory $\CL_{d-1}^{\rm even}$ of \eqref{eqnAWoddB} on the wall. See \cite{Queiroz_2016} for a related discussion in the Hamiltonian language.

By the same argument as above, the Berry number of $\CL_{2n-1}$ equals that of $\CL_{2n}$. Combining with the other inductive lemma, all of the Berry numbers of the Abanov-Wiegmann family are equal. Since the $D = 1$ case has Berry number 1 (the usual Berry number), all of the Berry numbers are 1.


Another important family is the Abanov-Wiegmann ``A" family, which realizes generalized Thouless pumps. In odd space dimensions $d$, the theories consist of $2^{(d-1)/2}$-many $2^{(d+1)/2}$ complex fermions $\psi_j$ transforming under the flavor group $Spin(d)$ with generators $\Gamma^a$ and the Lagrangian
\begin{equation}\label{eqnAWoddA}
\CL_d^{\rm odd} = i \bar \psi_j \gamma^\mu \partial_\mu \psi_j + M^0 \bar \psi_j \psi_j + i M^a \bar \psi_j \gamma^{0\cdots d}, \Gamma^a_{jk} \psi_k,
\end{equation}
where $\gamma^{0\cdots d}$ is the chirality operator. Meanwhile, in even space dimensions, the theories consist of $2^{d/2}$-many $2^{d/2}$-component complex fermions transforming under $Spin(d+1)$ with generators $\Gamma^a$ and the Lagrangian
\begin{equation}\label{eqnAWevenA}
\CL_d^{\rm even} = i \bar \psi_j \gamma^\mu \partial_\mu \psi_j + M^a \bar \psi_j \Gamma^a_{jk} \psi_k.
\end{equation}
One can repeat the derivations above to show that this series also dimensionally reduces along interfaces. We can use this to reduce all the way down to quantum mechanics, where we find a 1-component complex fermion, which depending on the sign of the chemical potential has either a neutral unoccupied or a charged occupied ground state. The generalized Thouless pump follows for the whole series from this simple observation.

\subsection{Anomalous Interfaces}

In our construction above, the interface was always trivially gapped, and the parameter space of the interface was reduced to a smaller-dimensional sphere $S^{d+1}$. This is unlike our prescription for studying boundary anomalies in Section \ref{secbulkboundary}, however they can be related. In particular, we will see that the boundary diabolical point is hiding ``inside" the $S^{d+1}$ we defined above.

Indeed, we can define an extension of the interface parameter space from $S^{d+1}$ to the interior ball $B^{d+2}$ by adding a ``radial" parameter which interpolates our great circular path with a straight line path. One of these paths passes through the massless point of the $d$-dimensional theory. In the free fermion theory, it is easy to show that the are gapless modes along the interface for any such path, generalizing the arguments of Jackiw-Rebbi. Even if we add interactions, however, because there is a higher Berry number on the boundary of this $B^{d+1}$, there is guaranteed to be a diabolical locus inside.

To directly relate the interfaces to the conclusions of Section \ref{secbulkboundary} however, we have to allow one of the end points to vary over the bulk parameter space, while the other is fixed at some value $p_0$, which represents the parameter value ``outside of the sample". In this case, our choice of boundary condition for each bulk parameter $p$ amounts to a choice of path $f_p(u)$ from $p$ to $p_0$. This defines a boundary condition for each $p \in S^{d+2}$. Because the sphere is not contractible, at least one of these paths $f_{p_*}(u)$ must pass through the massless point in the origin for some $u$ (or more generally through some diabolical locus analogous to the one defined above). The point $p_*$ thus defines a boundary diabolical point, as required by Section \ref{secbulkboundary}.

\section{Diabolical Points of High Codimension}\label{appnofurther}

We will argue by induction that a system in $d$ space dimensions has no protected diabolical points of codimension $m > d+3$. The argument is rigorous for quantum mechanical systems but only a sketch for higher dimensions.

For quantum mechanical systems, this can be argued as follows. First, by adding to the Hamiltonian a term proportional to the fermion parity, we can modify our family so that all ground states, including degenerate ones, have the same charge. By working only in this sector, we reduce the problem to eliminating a diabolical point of codimension $>3$ in an ordinary quantum mechanical system without any special symmetries.

Let us choose local coordinates on the parameter space so that the diabolical locus is at the origin.
We can continuously deform our family away from the diabolical locus (and without creating any other diabolical points) so that at some distance $R > 0$ from the origin the Hamiltonians are all (minus) projectors onto their unique ground state. The space of such projectors is a complex projective space $\mathbb{CP}^{l-1}$, where $l$ is the dimension of the Hilbert space. For any $m>2$, by adding spectator degrees of freedom, we can make $l$ big enough so that $\pi_m (\mathbb{CP}^{l-1}) = 0$.\footnote{Recall $\mathbb{CP}^{\infty}$ has homotopy type $K(\ZZZ,2)$. That is, all its homotopy groups except $\pi_2$ are zero, while $\pi_2$ is isomorphic to $\ZZZ$. In fact, by these same arguments we find $\CM_0$ is homotopic to $\mathbb{CP}^\infty$.} Once we do so, it is possible to extend our family of projectors to the interior of the ball of radius $R$. We can then linearly interpolate with our original family to rid ourselves of the diabolical points inside this ball, all the while not modifying the Hamiltonian on the sphere, so in particular no diabolical points leave the ball. We  are left with a completely nondegenerate phase diagram. 

Now suppose we have shown the result for dimensions smaller than $d$, and we have an $m$-parameter phase diagram of $d$ dimensional systems, $m > d + 3$, and a diabolical locus in some compact region. We choose some basepoint $p_0 \in S^{m-1}_R$ on a sphere of radius $R$ large enough to enclose the diabolical locus. We study interfaces from $p_0$ to other points $p \in S^{m-1}_R$ by choosing paths in our parameter space. This will introduce diabolical loci of codimension $m-1$ for the interface. By the inductive hypothesis, these interfaces may be smoothed so that there are no diabolical points amount the interface theories. Thinking of these interfaces as paths in $\CM_d$, we learn that our family on $S^{m-1}_R$ can be extended to a ball $B^m_R$. Linearly interpolating our original family with this one inside the ball we rid ourselves of any diabolical points while keeping the family unchanged for radii greater than $R$.

\end{document}